\documentstyle[12pt]{article}
 \def\be{\begin{eqnarray}}
 \newcommand{\bm}[1] {\mbox{\boldmath{$#1$}}}
 \def\ee{\end{eqnarray}}
\input psfig.sty
\begin{document} 
\pagestyle{empty}
\Huge{\noindent{Istituto\\Nazionale\\Fisica\\Nucleare}}

\vspace{-3.9cm}

\Large{\rightline{Sezione di ROMA}}
\normalsize{}
\rightline{Piazzale Aldo  Moro, 2}
\rightline{I-00185 Roma, Italy}

\vspace{0.65cm}

\rightline{INFN-1296/00}
\rightline{June 2000}

\vspace{1.cm}



\begin{center}{\large\bf Poincar\'e Covariant Current Operator and Elastic
Electron-Deuteron Scattering in the Front-form Hamiltonian Dynamics}  
\end{center}


\begin{center} F.M. Lev$^a$, E. Pace$^b$ and G. Salm\`e$^c$\end{center}

\noindent {\it $^a$Laboratory of Nuclear Problems, Joint Institute
for Nuclear Research, Dubna, Moscow region 141980, Russia}

\noindent{$^b$\it Dipartimento di Fisica, Universit\`a di Roma
"Tor Vergata", and Istituto Nazionale di Fisica Nucleare, Sezione
Tor Vergata, Via della Ricerca Scientifica 1, I-00133, Rome,
Italy}

\noindent{$^c$\it Istituto Nazionale di Fisica Nucleare, Sezione
di  Roma, P.le A. Moro 2, I-00185 Rome, Italy}


\begin{abstract}

The deuteron electromagnetic form factors, $A(Q^2)$ and $B(Q^2)$, and the tensor
polarization $T_{20}(Q^2)$, are unambiguously calculated within the front-form
relativistic Hamiltonian dynamics, by using a novel current, built up from
one-body terms, which fulfills Poincar\'e, parity and time reversal covariance,
together with Hermiticity and the continuity equation. A simultaneous
description of the experimental data for the three deuteron form factors is
achieved up to $Q^2 < 0.4
(GeV/c)^2$. At higher momentum transfer, different nucleon-nucleon interactions
strongly affect $A(Q^2)$, $B(Q^2)$, and $T_{20}(Q^2)$ and the effects of the
interactions can be related to $S$-state kinetic energy in the deuteron.
Different nucleon form factor models have huge effects on $A(Q^2)$, smaller
effects on $B(Q^2)$ and essentially none on $T_{20}(Q^2)$.

\end{abstract}

\vspace{4.5cm}
\hrule width5cm
\vspace{.2cm}
\noindent{\normalsize{Submitted to {\bf Phys. Rev. C} }}

\newpage
\pagestyle{plain}

\pagestyle{plain}
\section{Introduction}
\label{S1}

The deuteron is a fundamental system for our understanding of nuclear physics
and a challenge to our ability to describe nuclei as systems of interacting
nucleons with a well defined internal structure, without an explicit use of
their quark substructure. In particular elastic electron-deuteron scattering is 
a crucial test for deuteron models. 

There exists a wide literature (see, e.g.,
\cite{Kondra,Hummel,Gross,Carbo,Kapta,Phillips,Despla} and Refs. quoted therein) 
devoted to the investigation of deuteron electromagnetic (em) properties and in
particular to the accuracy of the one-body impulse approximation (IA) for the
current operator. It is usually believed that effects beyond IA, as
meson-exchange currents, $N\bar{N}$-pair creation terms ($Z$ graphs), isobar
configurations in the deuteron wavefunction, etc. are important for the
explanation of existing data. However, the contributions of these effects are
essentially model dependent  \cite{Schia} and, furthermore, obviously depend on
the reference frame (see, e.g., Refs. \cite{CCKP,LPS}). 

Since precise measurements of the deuteron
elastic form factors have been recently performed in a wide range of momentum
transfer, up to $Q^2 = -q_{\mu}^2 = 6 (GeV/c)^2$ for $A(Q^2)$
\cite{JLABA,JLABC}, theoretical models require a relativistic framework for a
reliable description of the available data. Furthermore, it has been recently
shown \cite{LPS3} that relativistic effects are relevant even for static
deuteron properties, as the magnetic and quadrupole moments.   

An essential requirement for relativistic approaches is the covariance of the
current operator with respect to Poincar\'e group transformations. This
requirement is non-trivial for systems of interacting particles, since some of
the generators are interaction dependent. 

A widely adopted relativistic framework for the study of deuteron em properties
is the front-form Hamiltonian dynamics (FFHD) {\em {with a finite number of
particles}} (see Refs. \cite{Dir,Ter} and Refs. \cite{Lev1,KP} for extensive
reviews), which gives the possibility to retain the large amount of successfull
phenomenology developed within the nonrelativistic approaches. Indeed, in the
FFHD seven, out of ten, Poincar\'e generators are interaction free, in
particular the boost generators, while $P^- = (P_0 - P_z)/\sqrt{2}$ ($P$ is the
total momentum of the system) and the rotations around the $x$ and $y$ axes
contain the dynamics. Only the two-nucleon state is usually considered and
the wavefunction of the system factorizes for any front-form boost in an
eigenfunction of the total momentum times an intrinsic wavefunction, depending
only on internal variables. Therefore, in the case of elastic $e-d$ scattering,
one can express the three deuteron em form factors, determined by three
independent matrix elements of the current, in terms of the deuteron internal
wave function and the elastic em nucleon form factors (f.f.). 
 
In the FFHD the em properties of the deuteron were usually studied in the
reference frame where $q^+=(q_0+q_z)/\sqrt{2} = 0$ ($q$ is the momentum
transfer) \cite{Kondra,CCKP,GrKo,Frank,Brod,FFS,Fred}. The one-body approximation
was used to define three matrix elements  of the $plus$ component of the current,
while the other matrix elements of the $plus$ component and the other components
of the current were properly defined in order to fulfill Poincar\'e covariance,
Hermiticity and current conservation. However, for spin-one systems, as the
deuteron, this procedure is not unique and gives rise to ambiguities in the
calculation of the form factors \cite{GrKo,KS}.

In Ref. \cite{LPS}, using a representation of the Poincar\'e group within FFHD,
we have shown that extended Poincar\'e covariance (i.e., Poincar\'e plus
parity, $\cal{P}$, and time reversal, $\cal{T}$, covariance) is fulfilled by the
current which has a one-body form in the Breit reference frame where the initial
and final momenta of the system are directed along the spin quantization axis
($\vec{q}_{\bot}=\vec{q} - q_z {\bm e}_z = 0$). Furthermore, we have shown that
Hermiticity and current conservation can be easily implemented. An important
feature of our approach is that it allows one to use the same definition for all
the matrix elements of the current, without ambiguities.
  
In a previous paper \cite{LPS3}, as a test of our current, we evaluated 
the deuteron form factors at $Q^2= 0$, namely the magnetic moment,
$\mu_d$, and the quadrupole moment, $Q_d$, of the deuteron, which are not
affected by the uncertainties in the knowledge of the neutron em form factors at
finite momentum transfers. The deuteron magnetic and quadrupole moments
represented a longstanding problem in nuclear physics. Indeed, theoretical
calculations were not able to accurately reproduce in a coherent approach the
experimental values for both quantities at the same time, although a variety of
approaches have been attempted, by changing the tensor content of the
nucleon-nucleon ($N-N$) interaction, or considering two-body current
contributions, both in non-relativistic and in relativistic frameworks
\cite{Lomon,AV18,Tjon,Kapta}. On the contrary, using our Poincar\'e covariant
current operator, this usual disagreement between theoretical and experimental
results was reduced to  $0.5\%$ for $\mu_d$ and to $2\%$ for $Q_d$ by using
interactions able to reproduce the experimental value of the deuteron asymptotic
normalization ratio $\eta = A_D/A_S$. Therefore the contributions from explicit
two-body currents or from isobar configurations in the deuteron wave function
should be relatively small at $Q^2= 0$.

 Encouraged by this result, in the present paper we study, within the
framework of FFHD and using our Poincar\'e covariant current operator, the
deuteron form factors at $Q^2 \neq 0$ and in particular the effects produced by
: i) different $N-N$ interactions, and ii) different nucleon form factors models.
We will also investigate the possibility to gain information from elastic $e-d$
scattering on the neutron em structure, and in particular on the neutron charge
form factor. Our preliminary results were already published in Ref. \cite{LPS2}.

The plan of the paper is the following:
in Sect. 2 the definition of our covariant current operator is recalled; in
Sect. 3 the elastic deuteron form factors are expressed in terms of the matrix
elements of the free current in the Breit frame; in Sect. 4 the
front-form deuteron wave function and the explicit expressions of the current
matrix elements in terms of the deuteron wave function are presented; in Sect.
5 our results on the dependence of deuteron form factors upon $N - N$
interactions and nucleon em form factors are discussed and, eventually, in Sect.
6 our conclusions are drawn.
 
\section{A covariant current operator within the front-form dynamics}
\label{S2}

In this section we give the essential lines for the definition of a current
which satisfies extended Poincar\'e covariance, Hermiticity, current
conservation, and charge normalization, to be applied to the calculation of
elastic em form factors.

Let us first consider the extended Poincar\'e covariance. If the current operator
$J^{\mu}(x)$ is defined in terms of $J^{\mu}(0)$ 
\begin{sloppypar}
\begin{equation}
J^{\mu}(x) = exp(\imath Px) J^{\mu}(0)exp(-\imath Px),
\label{31}
\end{equation}
then the Poincar\'e covariance of $J^{\mu}(x)$ takes place if
\begin{equation}
U(l)^{-1}J^{\mu}(0)U(l)=L(l)^{\mu}_{\nu}J^{\nu}(0) ,
\label{33}
\end{equation}

\noindent where $L(l)$ is the element of the Lorentz group corresponding to
$l\in SL(2,C)$
and $U(l)$ is the unitary representation operator corresponding to $l$
(see, e.g., \cite{Weinb}). 
\end{sloppypar} 

For systems of interacting particles the operator $U(l)$ in general does depend
on the interaction, and it is not trivial to build up a current which
satisfies Eq. (\ref{33}). Indeed, in order to fulfill this
requirement the current operator has to be interaction dependent. The key
property of our procedure \cite{LPS} for the definition of a Poincar\'e 
covariant current operator is the following spectral decomposition of the
current:

\begin{equation}
J^{\mu}(0)=\sum_{ij}\Pi_i J^{\mu}(0)\Pi_j
\label{39}
\end{equation}

\noindent In Eq. (\ref{39}) $\Pi_i$ is the orthogonal projector onto the 
subspace ${\cal H}_i\equiv \Pi_i{\cal H}$ corresponding to the
mass $M_i$, the spin $S_i$, and a definite parity, with ${\cal H}$ being
the space of states describing the interacting particle system. This
decomposition allows one to express the possible current operator dependence
on the interaction as a dependence on mass and spin of the
interacting particle system.

In the FFHD, the seven Poincar\'e generators belonging
to the subgroup which leaves invariant the hyperplane $x^+ = 0$ are kinematical.
Then, as already mentioned in the introduction, the state of a system,
$|P,\chi\rangle$, factorizes in a total momentum eigenstate, $|{\vec
P}_{\bot},P^{+}\rangle$, times an intrinsic eigenstate, $|\chi\rangle$ :

\begin{equation}
|P,\chi\rangle =  |{\vec P}_{\bot},P^{+}\rangle |\chi\rangle \; ,
\label{38}
\end{equation}
In Eq. (\ref{38})  $P^+= (P_0 + P_z)/\sqrt{2} = p_1^+ +...+ p_N^+$ and 
${\vec P}_{\bot} = (P_x,P_y) = {\vec p}_{1 \bot} + ... + {\vec p}_{N \bot} $
are the $plus$ and ${\bot}$ components of the total momentum, with
$p_1, ... , p_N$ the individual momenta of the particles in the system.  Because
of the decomposition of Eq. (\ref{39}), the operator $J^{\mu}(0)$ is fully
defined by the set of matrix elements between initial,
$| {\bm P}_{\bot},P_j^{+}\rangle$, and final, 
$| {\bm P}_{\bot}',P_i^{'+}\rangle$, total momentum eigenstates

\begin{equation}
J^{\mu}(P_i';P_j) \equiv \langle {\bm P}_{\bot}',P_i^{'+}|\Pi_iJ^{\mu}(0)\Pi_j|
{\bm P}_{\bot},P_j^{+}\rangle 
\label{42'}
\end{equation} 
The matrix elements between total momentum eigenstates, $J^{\mu}(P_i';P_j)$,
correspond to definite values of masses, spins and parity, and are operators in
the space  ${\cal H}_{int}$ of intrinsic states. Through proper unitary
transformations, the current operator $J^{\mu}(P_i',P_j)$ in any reference frame
can be defined in terms of the auxiliary current operators
\begin{eqnarray}
j^{\nu}(K{\bm e}_z;M_i,M_j)\equiv\langle {\bm K}_{i\bot}' 
= 0,K_i^{'+}|\Pi_iJ^{\mu}(0)\Pi_j| {\bm K}_{j\bot} = 0,K_j^{+}\rangle 
\label{43}
\end{eqnarray}
in the special Breit frame where the total three-momenta of the system in the
initial state, ${\bm K_j} = - K{\bm e}_z$, and in the final state, ${\bm K_i'}
= K{\bm e}_z$, are directed along the spin quantization axis, $z$.  In Eq.
(\ref{43}) the initial and final {\em plus} components of the total momentum are 
\begin{equation}
K^{+}_j = {1 \over \sqrt{2}} [(M_j + K^2)^{1/2} - K ], \quad
K^{'+}_i= \frac{1}{\sqrt{2}} [(M_i^2 + K^2)^{1/2} + K], 
\label{42''}
\end{equation}  
while $K=Q/2$,  and $q = K'_i - K_j$.
It has been shown \cite{LPS} that the operator $J^{\mu}(0)$
fulfills Eq. (\ref{33}), i.e. is Lorentz covariant, if the current operators
$j^{\nu}(K{\bm e}_z;M_i,M_j)$ in the above special Breit
frame are covariant with respect to rotations around the $z$ axis.

 Since in the front form the rotations around the $z$ axis are interaction
free, the continuous Lorentz transformations constrain the current
$j^{\mu}(K{\bm e}_z;M_i,M_j)$ for an interacting system in the same way as in
the non-interacting case. The same property holds for the covariance with 
respect to a reflection of the $y$ axis, ${\cal P}_y$, and with respect to the 
product
of parity and time reversal, $\theta$, which leave the light cone $x^+=0$
invariant, and therefore are kinematical. 
The full space reflection is the product of ${\cal P}_y$ and a dynamical
rotation around the $y$ axis by $\pi$, while ${\cal T}=\theta {\cal P}$,
and therefore parity and time reversal do not contain an interaction dependence
different from the one implied by rotations around $y$ axis. As a consequence,
the current operator satisfies ${\cal P}$ and ${\cal T}$ covariance, if it
satisfies Poincar\'e covariance and covariance with respect to $P_y$ and
$\theta$  \cite{LPS}.

In conclusion, since in our Breit frame the extended Poincar\'e covariance
constraints for the auxiliary operators are the same for a non-interacting and 
an interacting system, the extended Poincar\'e covariance is satisfied for an
interacting system by a current composed  by the sum of free,
one-body currents, viz. 
\begin{eqnarray}
j^{\mu}_{free}(K{\bm e}_z;M_i,M_j)\equiv\langle 0, K_i^{'+}|
\Pi_iJ^{\mu}_{free}(0)\Pi_j |  0,K_j^{+}\rangle  
\label{44} 
\end{eqnarray}
where $J^{\mu}_{free}(0) = \sum_{i=1}^{N} j_{free,i}^{\mu}$, with  N 
the number of constituents in the system.

 In the elastic case considered in this paper ($M_i=M_j=M$; $S_i=S_j=S$), 
the property of Hermiticity for the auxiliary operators reads as follows 
\begin{equation}  
j^{\mu}(-\bm{K};M,M)=j^{\mu}(\bm{K};M,M)^* ,
\label{91'}
\end{equation}
where $^*$ means the Hermitian conjugation in the internal space
${\cal{H}}_{int}$. For $|{\bm K}| = 0$ the property of Hermiticity reads
$j^{\mu}(0;M,M) = j^{\mu}(0;M,M)^*$, while
for $|{\bm K}|\neq 0$ it becomes a non-trivial constraint and is satisfied if 
\begin{eqnarray}
&&j^{\mu}(K{\bm e}_z;M,M)^*= \nonumber \\
&&L[r_x(-\pi)]^{\mu}_{\nu} D^{S}[r_x(-\pi)]
j^{\nu}(K{\bm e}_z;M,M) D^{S}[r_x(-\pi)])^{-1},
\label{60}
\end{eqnarray}
where  $D^s(u)$ is the matrix of the unitary irreducible
representation of the group $SU(2)$ with spin $s$, corresponding to
$u\in SU(2)$, and 
$r_x(-\pi)$ represents a rotation by $-\pi$ around the $x$ axis, i.e. 
$D^{S}[r_x(-\pi)] = exp(\imath \pi S_x)$ \cite{LPS}.

 Let $\Pi$ be the projector onto the  subspace of bound states $|\chi \rangle$
of mass $M$ and spin $S$, and let ${\cal{J}}^{\mu}(K\bm{e}_z;M,M)$ be a
current which fulfills extended Poincar\'e covariance.
Then a choice for the current compatible with the Hermiticity
condition, Eq. (\ref{60}), and with the extended Poincar\'e covariance is 
\cite{LPS}
\begin{eqnarray}
j^{\mu}(K\bm{e}_z;M,M) = \frac{1}{2}\{  {\cal{J}}^{\mu}
(K\bm{e}_z;M,M) +   \nonumber\\
L^{\mu}_{\nu}[r_x(-\pi)]~exp(\imath \pi S_x)~
\left[ {\cal{J}}^{\nu} (K\bm{e}_z;M,M) \right] ^*
~exp(-\imath \pi S_x)   \}.
\label{97}
\end{eqnarray}
The second term in Eq. (\ref{97}), which ensures Hermiticity,
introduces implicitely two-body terms in the current, because of the presence of
the $x$ component of the front-form spin operator, $S_x$. 

This current fulfills also the current conservation, which in the elastic case
 reads 
\begin{equation}
j^-(K\bm{e}_z;M,M) = j^+(K\bm{e}_z;M,M)
\label{minus'}
\end{equation}
Indeed, as shown in ref. \cite{LPS}, in the elastic case the extended Poincar\'e
covariance and Hermiticity imply Eq. (\ref{minus'}), i.e., impose current
conservation. 

In Eq. (\ref{97}) one has to choose a specific definition for the
operator  ${\cal{J}}^{\mu}(K\bm{e}_z;M,M)$. Unfortunately, one cannot simply
adopt Eq. (\ref{44}), because of the charge normalization condition, which
implies

\begin{eqnarray}
j^{+}(0;M,M) = \frac{1}{2} \{  {\cal{J}}^{+}
(0;M,M) +   
{\cal{J}}^{-} (0;M,M)   \} = \sqrt{2} e M \Pi.
\label{90}
\end{eqnarray}
where $e$ is the total electric charge of the system. Indeed, while the charge 
normalization condition is fulfilled by $j^{+}_{free}(0;M,M)$, 
Eq. (\ref{90}) is not satisfied by $\frac{1}{2}(j^{+}_{free}(0;M,M) +
j^{-}_{free}(0;M,M))$. However, a possible choice is the following one: 
\begin{eqnarray} {\cal{J}}^{+}(K\bm{e}_z;M,M) =
\langle 0, K^{'+}|\Pi J_{free}^{+}(0) \Pi| 0, K^{+} \rangle \nonumber\\
\bm{{\cal{J}}}_{\perp}(K\bm{e}_z;M,M) =
\langle 0, K^{'+}|\Pi \bm{J}_{\perp free}(0) \Pi| 0, K^{+} \rangle \nonumber\\
{\cal{J}}^{-}(K\bm{e}_z;M,M)={\cal{J}}^{+}(K\bm{e}_z;M,M).
\label{95}
\end{eqnarray}
The previous definition of the "$-$" component of ${\cal{J}}^{\mu}$ is essential
for the proper charge normalization of $j^{\mu}(K\bm{e}_z;M,M)$, because of the
second term in Eqs. (\ref{97},\ref{90}).

In the elastic case, only $2S+1$ non-zero matrix elements of the em current
defined by Eqs. (\ref{97},\ref{95}) are independent, corresponding to the
$2S+1$ elastic form factors. The independent matrix elements can be chosen as
the diagonal matrix elements of $j^+$  with $S_z\geq 0$ and the matrix elements 
$\langle M S S_z| j_x(K\bm{e}_z;M,M) | M S S_z-1 \rangle$ of $j_x$ with 
$S_z\geq + 1/2$ \cite{LPS}. Obviously, any other choice of the independent
matrix elements to be used in the calculation of the elastic form factors is
completely equivalent, i.e. it will yield exactly the same results. One can
immediately obtain that

\begin{eqnarray}
&&\langle M S S_z| j^+(K\bm{e}_z;M,M) | M S S_z \rangle =
 \langle M S S_z| {\cal J}^+(K\bm{e}_z;M,M) | M S S_z \rangle,
 \label{106} \\
&& \langle M S S_z| j_x(K\bm{e}_z;M,M) | M S S'_z \rangle = {1 \over 2}
[ \langle M S S_z| {\cal J}_x(K\bm{e}_z;M,M)
| M S S'_z \rangle -  \nonumber\\
&&\langle M S S'_z| {\cal J}_x(K\bm{e}_z;M,M) | M S S_z \rangle ]
\label{107}
\end{eqnarray}
and therefore the elastic form factors can be evaluated in terms of the matrix
elements of the free current only. It has to be noted that the matrix elements of
both $j^+$ and $j_x$ have been shown to be real \cite{LPS}.

In the deuteron case, since $S=1$, three matrix elements of the current are
needed.

\section{Deuteron electromagnetic form factors}
\label{S3}

The form factors $A(Q^2)$ and $B(Q^2)$, which appear in the
unpolarized cross section, and the tensor polarization, $T_{20}(Q^2)$, can be
expressed in terms of the charge, $G_C(Q^2)$, quadrupole,
$G_Q(Q^2)$, and magnetic, $G_M(Q^2)$, elastic 
form factors :

\begin{eqnarray}
&&A(Q^2) = G_C^2+ \frac{8}{9} \tau ^2 G_Q^2+ \frac{2}{3} \tau G_M^2 \nonumber \\
&&B(Q^2) = \frac{4}{3} \tau ( 1 + \tau ) G_M^2 \nonumber \\
&&T_{20}(Q^2) = - \tau \frac{\sqrt{2}}{3} 
\frac{ [ \frac{4}{3} \tau G_Q^2 + 4 G_Q G_C + f G_M^2 ] }{A + B \tan ^2(\theta /
2)} 
\label{98'}
\end{eqnarray}
where $\tau = Q^2/(4 m^2_d)$, $Q^2=-q^2_{\mu}$, $m_d$ is the deuteron mass
and $f = 1/2 + (1 + \tau ) \tan ^2 (\theta / 2)$, with the
following normalization for the form factors: $G_C(0) = 1$, $G_Q(0) = m_d^2 Q_d$,
and $G_M(0) = \mu _d m_d/m_p$ ($m_p$ is the proton mass).

For the deuteron, the matrix elements of the current are related to the 
form factors $G_C(Q^2)$, $G_M(Q^2)$, $G_Q(Q^2)$ by the following general 
expression of the macroscopic current for spin $1$ systems (as
the deuteron) \cite{Glaser}  
\begin{eqnarray}
&&j^{\mu}_{S_z',S_z} = \langle m_d 1 S_z'| j^{\mu}(K\vec{e}_z, m_d, m_d) | m_d 1
S_z \rangle = e e^{'*\alpha}_{S_z'}e^{\beta}_{S_z} \left\{ (P + P')^{\mu} 
\left[ - (G_C - \frac{2}{3} \tau G_Q) g_{\alpha \beta} \right. \right. \nonumber
\\  &&\left. \left.
-  \zeta ^2 [G_C-(1 + \frac{2}{3} \tau )G_Q-G_M] q_{\alpha} q_{\beta}
\right] +  G_M \left( g_{\alpha}^{\mu}q_{\beta} - g_{\beta}^{\mu}q_{\alpha}
\right) \right\}
\label{100} 
\end{eqnarray}
where $| m_d 1 S_z \rangle $ is the deuteron intrinsic eigenstate, 
$g_{\alpha \beta}$ the metric tensor,
$e_{S_z}$ and $e'_{S_z'}$ are the initial and final deuteron polarization
vectors, respectively, (see Appendix A) and $\zeta^{-1} = \sqrt{2} m_d \sqrt{1
+ \tau}$.

In FFHD, hadron form factors are often calculated in
the reference frame where $q^+=0$. If $\lambda$ and $\lambda'$
are the helicities in the initial and final states, respectively, and
$I_{\lambda'\lambda}=\langle \lambda'|J^+(0)|\lambda\rangle $,
then, because of Hermiticity, ${\cal P}$ and ${\cal T}$ covariance, and 
covariance for rotations about the $z$ axis, all the matrix
elements $I_{\lambda'\lambda}$ for the deuteron can be expressed in terms of
$I_{11}$, $I_{00}$, $I_{10}$ and $I_{1,-1}$. As shown, e.g., in Refs.
\cite{GrKo,CCKP}, the following constraint, usually called "angular
condition", must be fulfilled in the $q^+=0$ frame, viz. 
\begin{equation}
(1+2\tau)I_{11}-I_{00}-(8\tau)^{1/2}I_{10}+I_{1,-1} = 0 .
\label{1}
\end{equation}
However, this constraint, which is related to the rotational covariance of the
current, is not satisfied if the matrix elements $I_{\lambda'\lambda}$  are
calculated with the free operator, $J_{free}^+(0)$ in the $q^+=0$ frame. 
Then, three out of the four
matrix elements are usually defined through the free operator, while the fourth
one is defined by Eq. (\ref{1}). However, different choices of the
three matrix elements to be calculated by the free operator are
possible and therefore different prescriptions can be used to calculate the
three physical form factors. As a consequence, within this approach there is a
large ambiguity in the theoretical results (see, e.g., 
\cite{CCKP,GrKo,Frank,Brod,FFS,KS}), and,
furthermore, different definitions are used for different matrix elements of the
current.

 A relevant result of our approach is that, using in the left hand side 
of Eq. (\ref{100}) the microscopic current defined by Eqs. (\ref{97},\ref{95}),
the extraction of elastic em form factors is no more plagued by the
ambiguities, which are present when the free current is used in the reference
frame where $q^+=0$. Indeed, using our current operator, it turns out that
only three matrix elements $j^{\mu}_{S_z',S_z}$ are independent, corresponding 
to the three elastic em form factors. For instance, one can consider the matrix
elements $j^{+}_{0,0}$, $j^{+}_{1,1}$, $j^{x}_{1,0}$, which have been shown to
be real  \cite{LPS}. On
the contrary, using the one-body current in the $q^+=0$ frame, one has four
independent matrix elements \cite{GrKo}.

The form factors $G_C$, $G_M$, and $G_Q$ can be easily obtained from the matrix
elements of the current in our Breit frame, since from Eq. (\ref{100}) one has
\begin{eqnarray}
&&\langle m_d 1 1| j^+(K\bm{e}_z;m_d,m_d) | m_d 1 1 \rangle = 
\zeta^{-1} \left[ G_C - \frac{2}{3} \tau G_Q \right]   \nonumber \\
&&\langle m_d 1 0| j^+(K\bm{e}_z;m_d,m_d) | m_d 1 0 \rangle = 
\zeta^{-1} \left[ G_C + \frac{4}{3} \tau G_Q \right]   \nonumber \\
&&\langle m_d 1 1| j_x(K\bm{e}_z;m_d,m_d) | m_d 1 0 \rangle = 
\zeta^{-1} \tau ^{\frac{1}{2}} G_M 
\label{2}
\end{eqnarray}

By means of Eq. (\ref{2}) and using the properties (\ref{106}), (\ref{107}) of
the matrix elements $j^{\mu}_{S_z',S_z}$, the form factors
$G_C$, $G_M$, and $G_Q$ can be expressed in terms of the matrix elements
${\cal{J}}^{+}_{S_z,S_z} =  \langle m_d 1 S_z| {\cal{J}}^{+}(K\vec{e}_z,
m_d,m_d) | m_d 1 S_z \rangle$ and  ${\cal{J}}^{x}_{S_z',S_z} =  \langle m_d 1
S_z'| {\cal{J}}_{x}(K\vec{e}_z, m_d,m_d) | m_d 1 S_z \rangle$, i.e. in terms of
the matrix elements of the free current, calculated in the Breit
frame where the momentum transfer is along the spin quantization axis, $z$
\cite{LPS2}. One obtains        
\begin{eqnarray} 
G_C = (2 {\cal{J}}^{+}_{1,1} + {\cal{J}}^{+}_{0,0}) \zeta /3,  \quad
G_M = ({\cal{J}}^{x}_{1,0} - {\cal{J}}^{x}_{0,1}) \zeta / (2 \sqrt{\tau}), \quad
G_Q = ({\cal{J}}^{+}_{0,0} - {\cal{J}}^{+}_{1,1})  \zeta / (2\tau). 
\label{98'''}
\end{eqnarray}

\noindent Then, the deuteron magnetic moment, in nuclear magnetons, is given
by    
\begin{equation}
\mu _d = \frac{m_p}{(\sqrt{2} m_d)} \lim_{Q \rightarrow 0}   \frac{1}{ Q}
[{\cal{J}}^{x}_{1,0} - {\cal{J}}^{x}_{0,1}]  ,
\label{99'}
\end{equation}
while the deuteron quadrupole moment is
\begin{eqnarray}
 Q_d = \frac{\sqrt{2}}{ m_d} \lim_{Q \rightarrow 0}  \frac{1}{ Q^2}  
[ {\cal{J}}^{+}_{0,0} - {\cal{J}}^{+}_{1,1} ] . 
\label{101}
\end{eqnarray}
We stress that, as was shown in \cite{Kondra}, using the free current in the
frame where $q^+=0$, in the limit $Q^2 \rightarrow 0$ the angular condition is
satisfied at the first order in $Q$, but it is violated at the second order.
Therefore the angular condition is not a problem for the calculation of $\mu_d$,
while the quadrupole moment is not uniquely determined within that approach.

 From Eqs. (\ref{98'},\ref{98'''}) it is straightforward to obtain the
expressions for the elastic structure functions $A(Q^2)$, $B(Q^2)$  and for the
tensor polarization $T_{20}(Q^2)$ in terms of the matrix elements of the free
current ${\cal{J}}^{+}_{S_z,S_z}$ and ${\cal{J}}^{x}_{S_z',S_z}$:
\begin{eqnarray}
&&A(Q^2) = \frac{{\zeta}^2}{3} [ ({\cal{J}}^{+}_{0,0})^2 +
2 ({\cal{J}}^{+}_{1,1})^2 +  ({\cal{J}}^{x}_{1,0} - {\cal{J}}^{x}_{0,1})^2 / 2 ]
\nonumber \\ 
&&B(Q^2) = \frac{1}{6 m_d^2} ({\cal{J}}^{x}_{1,0} - {\cal{J}}^{x}_{0,1})^2 
\nonumber \\ 
&&T_{20}(Q^2) = - {\zeta}^2 \frac{\sqrt{2}}{3}  
\frac{ [ ({\cal{J}}^{+}_{0,0})^2 - ({\cal{J}}^{+}_{1,1})^2 + 
f ({\cal{J}}^{x}_{1,0} - {\cal{J}}^{x}_{0,1})^2 / 4 ] }{A + B \tan ^2(\theta /
2)}     
\label{98''}  
\end{eqnarray}

\section{Deuteron front-form wave function and matrix elements of the current
operator} 
\label{S4}

We consider the deuteron as a system of two different, interacting particles
with the same mass, $m = (m_p + m_n)/2$ ($m_n$ is the neutron mass), and spin
$1/2$. For a system of $N$ particles with four-momenta $p_i$
$(i=1,2,...,N)$, FFHD internal variables ${\bm k}_1$, ..., ${\bm k}_N$ can
be defined, such that $\sum_{i=1}^{N} {\bm k}_i = 0$ . The intrinsic
three-momentum ${\bm k}_i$ is the spatial part of the four-vector 
\begin{equation}
k_i=L[\beta(G)]^{-1}p_i,
\label{15}
\end{equation}
where  $G=P_0/M_0$ is the four-velocity, and $P_0=p_1 + ... + p_N$ the total
four-momentum of a system of free particles, with $M_0=|P_0|\equiv
|P^2_0|^{1/2}$.   \noindent The matrix $\beta(G)\in$ SL(2,C) (see Appendix B)
represents  a front-form boost. The action of the boost
$L[\beta(G)]^{-1}$ is such that $P^{'}_0=L[\beta(G)]^{-1}P_0
\equiv[{\bm P}^{'}_{0 \bot}=0, P_0^{'+}=M_0,P_0^{'-}=M_0$].

Then the wave function for the deuteron internal state $| m_d 1 S_z \rangle 
\equiv |\chi_{1,S_z}\rangle$
can be written as follows \cite{LPS1}
\begin{equation}
\chi_{1,S_z}({\vec k}_{\bot},\xi,\sigma_1,\sigma_2) =
\langle {\vec k},\sigma_1,\sigma_2|\chi_{1,S_z}\rangle =
\langle {\vec k},\sigma_1,\sigma_2|R^{-1} |\Psi_d\rangle
\omega(k)^{1/2}, 
\label{5}
\end{equation}
where
$\xi = p_1^+ / P^+$, and ${\vec k}_{\bot}={\vec p}_{1\bot}- 
\xi{\vec P}_{\bot}$. The internal three-momentum is ${\vec k}=({\vec
k}_{\bot},k_z)$, where $k_z = (2 \xi - 1) \omega(k)$, $\omega(k) = (m^2 + {\vec
k}^2)^{1/2}$, and $k = |{\vec k}|$.  It can be easily shown that $M_0 = 2
\omega(k)$. The normalization of 
$\langle {\vec k},\sigma_1,\sigma_2|\chi_{1,S_z}\rangle$ is such that
\begin{equation} 
\sum_{\sigma_1,\sigma_2} \int\nolimits
| \langle {\vec k},\sigma_1,\sigma_2|\chi_{1,S_z}\rangle |^2 
\frac{d{\vec k}}{(2\pi)^3 \omega (k)}  = 1
\label{5'}
\end{equation}

The matrix $R$ is given by 
\begin{equation}
R = v({\vec k, \vec s_1}) v(-{\vec k, \vec s_2})
\label{15'}
\end{equation}
where $v({\vec k, \vec s})$ is the Melosh matrix \cite{Mel,Ter} 
\begin{equation}
v({\vec k, \vec s}) = \frac{\omega(k) + m + k_z +
\imath (\hat{\sigma}_x k_y - \hat{\sigma}_y k_x) }
{[2(\omega(k)+m)(\omega(k)+k_z)]^{1/2}} , \label{27}
\end{equation}
while $\vec s_1$, and $\vec s_2$ are the usual nucleon spin operators,
$\sigma_1$ and $\sigma_2$ the eigenvalues of $s_{1z}$ and $s_{2z}$,
respectively, and $\hat{\sigma}_i$ the Pauli
matrix operators. 
The generalized Melosh matrix can also be written as
\begin{equation}
v({-\bm k},{\bm s}) = exp(\frac{\imath}{2}\varphi {\bm n} 
{\bm {\hat{\sigma}}}),  
\label{12} 
\end{equation}
with ${\bm n} = ({{\bm e}_z}\wedge \bm k)/k_{\bot}$,
by defining the angle $\varphi$
\begin{equation}
\varphi = 2arctan \frac{k_{\bot}}{\omega(k)+m-k_z},
\label{13}
\end{equation}
The angle $\varphi$ will be used in the Appendix for the calculation of the
deuteron quadrupole moment. 

The wave function for the deuteron internal state obeys the mass equation
\begin{equation}
M^2 \chi_{1,S_z}({\vec k}_{\bot},\xi,\sigma_1,\sigma_2) =
m_d^2 \chi_{1,S_z}({\vec k}_{\bot},\xi,\sigma_1,\sigma_2) 
\label{14'}
\end{equation}
while the wave function $\Psi_d$ in Eq. (\ref{5}) is the
usual solution of the "nonrelativistic" Schroedinger equation.
Indeed, if in the front-form dynamics the mass operator ${\tilde M}$ for the
function $\Psi_d$ is
defined by ${\tilde M}^2 = R M^2 R^{-1} = M_0^2 + V$ with $V$ the interaction
operator, then the mass equation ${\tilde M}^2\Psi_d = m_d^2\Psi_d$
has the same form as the "nonrelativistic" Schroedinger equation in momentum
representation \cite{Coester,Ter}:

\begin{equation}
(\frac{{\vec k}^2}{m}+ {\cal V} )\Psi_d({\vec
k},\sigma_1,\sigma_2) = E_d\Psi_d({\vec k},\sigma_1,\sigma_2)
\label{13'}
\end{equation}

\noindent where

\begin{equation}
{\cal V} = V/4m,\quad E_d=(m_d^2-4m^2)/4m = \epsilon_d + \epsilon_d^2/(4m)
\label{14}
\end{equation}

\noindent with $m_d = 2 m + \epsilon_d$. Therefore the eigenvalue $E_d$ of Eq.
(\ref{13'}) can be identified with the deuteron energy $\epsilon_d$, if the small
quantity $\epsilon_d^2/(4m)$ is disregarded. It has to be noted that, in
the case of the $N-N$ interactions of the Nijmegen group \cite{Nij}, $E_d$
is directly linked through Eq. (\ref{14}) to the deuteron energy
$\epsilon_d$ used in their fits. For the continuous part of the two-nucleon
spectrum the mass equation is identical to the "nonrelativistic" Schroedinger
equation in momentum representation \cite{Lev1}. Therefore the operator
$\cal V$ has to satisfy the same constraints of the potential as in
nonrelativistic quantum mechanics and can be chosen to have any of the forms
usually employed for the $N - N$ interaction in nonrelativistic nuclear physics.

Since the wave function $\Psi_d$ is an eigenstate of the standard nonrelativistic
spin operator \cite{CCKP,Lev1,KP}
\begin{equation}
\vec S_{nr} = \vec {\em{l}}(\vec k) + \vec s_1 + \vec s_2
\label{18}
\end{equation}
where $\vec {\em{l}}(\vec k)$ is the usual orbital angular momentum, the
Clebsh-Gordan coupling coefficients can be used. Then the internal deuteron wave
function $\chi_{1,S_z}({\vec k},\sigma_1,\sigma_2)$ with polarization vector
${\vec e_{S_z}}$ (see Appendix A), is given by (cf. \cite{CCKP})  
\begin{eqnarray} 
&&\langle {\vec k}, \sigma_1,\sigma_2
| \chi_{1,S_z} \rangle = (2\pi)^{3/2} \sqrt{\omega (k)/2}
\sum_{\sigma_1'\sigma_2'}  
\left[ v({\vec k, \vec s_1})^{-1} \right]_{\sigma_1,
\sigma'_1}  
\left[ v(-{\vec k, \vec s_2})^{-1} \right]_{\sigma_2,\sigma'_2} \cdot
\nonumber\\ 
&& \left[ \varphi_0(k)\delta_{ij} -
\frac{1}{\sqrt{2}}(\delta_{ij}- \frac{3k_ik_j}{k^2})\varphi_2(k) \right] 
\left[ \hat{\sigma}_i \hat{\sigma}_y \right]_{\sigma'_1,\sigma'_2} 
{(\vec {e}_{S_z})}_j = \nonumber\\ 
&&2 \sqrt{\pi^{3}\omega (k)} {(\vec {e}_{S_z})}_j 
\left[ \chi_0(k)\delta_{ij} + \frac{3k_ik_j}{ \sqrt{2} k^2})\varphi_2(k) \right] 
\left[ v({\vec k, \vec s_1})^{-1} \hat{\sigma}_i
\hat{\sigma}_y v(-{\vec k, \vec s_2})^{*} \right]_{\sigma_1,\sigma_2} 
\label{26}
\end{eqnarray}
where a sum over the repeated indices $i,j=1,2,3$ is assumed and
$\chi_0(k) = \varphi_0(k) - (1/\sqrt{2})\varphi_2(k) $.
The wave functions $\varphi_0(k)$ and
$\varphi_2(k)$ coincide with the nonrelativistic $S$ and $D$ waves in momentum
representation \cite{Coester}. The normalization of $\varphi_0(k)$ and
$\varphi_2(k)$ is such that $\int\nolimits
[\varphi_0(k)^2+\varphi_2(k)^2]d{\vec k}=1$.
For the calculation of the matrix elements of the current it
will be useful to put the internal deuteron wavefunction in a
more compact form
\begin{eqnarray}  
&&\chi_{1,S_z}({\vec k},\sigma_1,\sigma_2) = 
2 \sqrt{\pi^{3} \omega (k)} {(\vec {e}_{S_z})}_j
F_{ij}({\bm k}) 
\left[ C_{i}({\bm k}) + \imath {\bm {\hat {\sigma}}} \cdot {\bm D}_{i}({\bm k})
\right]_{\sigma_1,\sigma_2}
\label{260}
\end{eqnarray}
where
\begin{eqnarray}
F_{ij}({\bm k}) =  \left[ \chi_0(k)\delta_{ij} +
\frac{3k_{i}k_{j}}{\sqrt{2}k^{2}} \varphi_2(k) \right] 
\label{60'}
\end{eqnarray}
and
\begin{eqnarray}
C_{i}({\bm k}) + \imath {\bm {\hat{\sigma}}} \cdot {\bm D}_{i}({\bm k}) =
v({\bm k},\vec s_1)^{-1} \hat{\sigma}_{i} \hat{\sigma}_y 
[v(-{\bm k},\vec s_2)]^* , \quad i = 1, 2, 3.
\label{61}
\end{eqnarray}
In this paper the matrices $C_{i}({\bm k}) + \imath {\bm {\hat{\sigma}}} 
\cdot {\bm D}_{i}({\bm k})$ will be called "generalized Melosh matrices for the
deuteron wave function". Explicit expressions for the real quantities
$C_{i}({\bm k}), {\bm D}_{i}({\bm k})$  can be found in Appendix C. 

The matrix elements ${\cal{J}}^{\mu}_{S_z',S_z}$ can be easily calculated, by
using the action of the free current on a two-body state 
$| \vec{P}_{\bot}, P^+ \rangle | \chi_{S,S_z} \rangle$ \cite{LPS1}:  
\begin{eqnarray} 
&&\langle p_1', p_2'; \sigma_1',\sigma_2' |
J_{free}^{\mu}(0) | \vec{P}_{\bot}, P^+  \rangle | \chi_{S,S_z} \rangle =
\sum_{\sigma_1} \bar{w}(p_1',\sigma_1') \cdot \nonumber\\
&&\left\{ 2m \left[ f_e^{is}((p_1'-p_1)^2) -    
 f_m^{is}((p_1'-p_1)^2) \right] \frac{(p_1+p_1')^{\mu}}
{(p_1+p_1')^2} + f_m^{is}((p_1'-p_1)^2) \gamma^{\mu} \right\} \cdot \nonumber\\
&&w(p_1,\sigma_1) 
\langle {\vec k},\sigma_1,\sigma_2' | \chi_{S,S_z} \rangle  \frac{1} {\xi}  
\label{18'} 
\end{eqnarray}
where, in our case,

\begin{equation}
J_{free}^{\mu}(0) = J_p^{\mu}(0) + J_n^{\mu}(0) . 
\label{94}  
\end{equation}  
In Eq. (\ref{18'}) $w(p,\sigma)$ is the front-form Dirac spinor \cite{LPS1}
(see Appendix B), while $f_e^{is} = f_e^p + f_e^n$ and $f_m^{is} = f_m^p +
f_m^n$ are the isoscalar electric and magnetic Sachs form factors of the
nucleon.

An explicit calculation, with the help of the matrix elements of the
$\gamma$ matrices between front-form Dirac spinors reported in Appendix B, shows
that, as a consequence of Eqs. (\ref{18'},\ref{94}),

\begin{eqnarray}
&&\langle \chi_{1,S_z} |{\cal{J}}^{+}(K\vec{e}_z,m_d,m_d)| \chi_{1,S_z}\rangle =
\langle \chi_{1,S_z} | \langle 0, K^{'+}| J_{free}^{+}(0) | 0, K^{+} \rangle
| \chi_{1,S_z} \rangle =  \nonumber\\
\nonumber\\
&&\sqrt{2} m_d \sum_{\sigma_1,\sigma_1'\sigma_2}\int\nolimits
\chi_{1,S_z}({\vec k}',\sigma_1',\sigma_2)^*
\left\{ \frac{am(f_e^{is} - f_m^{is}) [am + \imath b({\hat{\sigma} k)}_{\bot}]}
{a^2m^2+ b^2{\vec k}_{\bot}^2}+f_m^{is} \right\}_{\sigma_1'\sigma_1} \cdot
\nonumber\\ 
&&\chi_{1,S_z}({\vec k},\sigma_1,\sigma_2) (\xi \xi')^{1/2}
\frac{d{\vec k}'}{(2\pi)^3\omega (k')\xi} \\ 
\label{21}
\nonumber\\  
\nonumber\\  
&&\langle \chi_{1,S_z'} |{\cal{J}}_{x}(K\vec{e}_z,m_d,m_d)|\chi_{1,S_z}\rangle =
\langle \chi_{1,S_z'} | \langle 0, K^{'+}| J_{free}^{x}(0) | 0, K^{+} \rangle
| \chi_{1,S_z}\rangle = \nonumber\\
\nonumber\\
&&\sum_{\sigma_1,\sigma_1'\sigma_2}\int\nolimits
\chi_{1,S_z'}({\vec k}',\sigma_1',\sigma_2)^*
\left\{ \frac{4mk_x(f_e^{is}-f_m^{is}) [am + \imath b({\hat{\sigma} k})_{\bot}]}
{a^2m^2 + b^2{\vec k}_{\bot}^2} + 
f_m^{is} [ak_x + \imath b(m {\hat{\sigma}_y} + k_y {\hat{\sigma}_z)}]
\right\}_{\sigma_1'\sigma_1}\nonumber\\ 
&&\chi_{1,S_z}({\vec k},\sigma_1,\sigma_2) \frac{d{\vec k}'}{(2\pi)^3\omega
(k')\xi} 
\label{22}
\end{eqnarray}
where 
\begin{eqnarray}
&& a = \left[ \frac{K^{'+} \, \xi'}{K^{+} \, \xi} \right]^{1/2} +
\left[ \frac{K^{+} \, \xi}{K^{'+} \, \xi'} \right]^{1/2}, \quad
b = \left[ \frac{K^{'+} \, \xi'}{K^{+} \, \xi} \right]^{1/2} -
\left[ \frac{K^{+} \, \xi}{K^{'+} \, \xi'} \right]^{1/2} 
\label{4}
\end{eqnarray}
and the form factors $f_e^{is}$ and $f_m^{is}$ are functions of $(p_1'-p_1)^2$.
In our Breit reference frame, where $\vec{K}_{\bot}=0$ and $\vec{q}_{\bot}=0$,
the relations between the internal (${\vec k}_{\bot}$, $k_z$) and individual
nucleon variables, in the initial, $\chi_{1,S_z}({\vec k},\sigma_1,\sigma_2)$,
and final, $\chi_{1,S_z'}({\vec k}',\sigma_1',\sigma_2')$, 
wave functions are given by   
\begin{eqnarray} 
&&{\vec p}_{1\bot}={\vec p}_{1\bot}'={\vec k}_{\bot} = {\vec
k}_{\bot}', \quad p_1^+=\xi K^{+},\quad k_z = \omega(k) (2 \xi - 1), 
\quad k'_z = \omega(k') (2 \xi' - 1),\nonumber\\
\nonumber\\
&&\xi' = \frac{p'^{+}_1} { K^{'+}} = 1 + (\xi - 1) \frac{K^{+}} { K^{'+}} = 
\frac {\xi [ \sqrt{m_d^2 + K^2} - K ] + 2K } {\sqrt{m_d^2 + K^2} + K} =
\frac{\xi [\sqrt{1 + \kappa^2} - \kappa] + 2\kappa}{\sqrt{1 + \kappa^2} + \kappa}
\,.   
\label{19}
\end{eqnarray} 
with $\kappa = K/m_d$.
It is important to note that nucleon form factors cannot be
factorized out in the current matrix elements, since from Eq. (\ref{19}) one has
\begin{eqnarray}
(p_1' - p_1)^2 = -4\tau (m^2+{\vec k}_{\bot}^2)/(\xi\xi') \neq -Q^2.
\label{119}
\end{eqnarray} 
 By using the expression (\ref{26}) of the deuteron wave function, a direct
calculation shows that 
\begin{eqnarray}
&&\langle \chi_{1,S_z} |{\cal{J}}^{+}(K\vec{e}_z,m_d,m_d)| \chi_{1,S_z}\rangle =
\sqrt{2}m_d {(\vec {e}_{S_z})}_{j'}^{*} {(\vec {e}_{S_z})}_{j}
\int\nolimits \left[ \frac{\omega (k)\xi'}{\omega (k')\xi} \right] ^{1/2} \cdot
\nonumber\\ 
&& \left[ \chi_0(k')\delta_{i'j'}+
\frac{3k_{i'}'k_{j'}'}{\sqrt{2}{k'}^2}
\varphi_2(k') \right] 
\left[ \chi_0(k)\delta_{ij}+
\frac{3k_{i}k_{j}}{\sqrt{2}k^{2}}
\varphi_2(k) \right] \cdot
\nonumber\\
&&\frac{1}{2}Tr \left\{ [v(-{\bm k}',\vec s_2)]^{T}{\hat{\sigma}}_y
{\hat{\sigma}}_{i'}v({\bm k'}, \vec s_1)  
\left[ \frac{am(f_e^{is} - f_m^{is}) 
\left[ am + \imath b({\hat{\sigma}} k)_{\bot} \right] }
{a^2m^2+ b^2k_{\bot}^2} + f_m^{is} \right] \right. \cdot
\nonumber\\
&&\left. \frac{}{} v({\bm k},\vec s_1)^{-1}{\hat{\sigma}}_{i} {\hat{\sigma}}_y
[v(-{\bm k},\vec s_2)]^* \right\}   d{\bm k}' 
\label{8}
\end{eqnarray}
\begin{eqnarray}
&& \langle \chi_{1,S_z'} |{\cal{J}}_{x}(K\vec{e}_z,m_d,m_d)|\chi_{1,S_z}\rangle =
{(\vec {e}_{S_z'})}_{j'}^{*} {(\vec {e}_{S_z})}_{j}
\int\nolimits
\left[ \frac{\omega (k)}{\omega (k')} \right] ^{1/2} \cdot
\nonumber\\
&& \left[ \chi_0(k')\delta_{i'j'} + \frac{3k_{i'}'k_{j'}'}
{\sqrt{2}{k'}^2}\varphi_2(k') \right] 
\left[ \chi_0(k)\delta_{ij}+
\frac{3k_{i}k_{j}}{\sqrt{2}k^{2}}
\varphi_2(k) \right] \cdot
\nonumber\\
&&\frac{1}{2}Tr \left\{ [v(-{\bm k}',\vec s_2)]^{T}
{\hat{\sigma}}_y {\hat{\sigma}}_{i'} v({\bm k}', \vec s_1)
\left\{ \frac{4mk_x(f_e^{is} - f_m^{is}) [am + \imath b({\hat{\sigma}}
k)_{\bot}]} {a^2m^2 + b^2k_{\bot}^2} \; + \right. \right.
\nonumber\\
&& \left. \left. \frac{}{} f_m^{is} [ak_x + \imath b(m {\hat{\sigma}}_y
+ k_y {\hat{\sigma}}_z)] \right\} v({\bm k},\vec
s_1)^{-1}{\hat{\sigma_{i}}}{\hat{\sigma}}_y [v(-{\bm k},\vec s_2)]^* \right\}
\frac{d{\bm k}'}{\xi} 
\label{9}
\end{eqnarray}
where the superscript $T$ on a matrix
indicates the transposition of the matrix and a sum over the repeated indices
$i, j, i', j'$ is understood. By means of the matrices $C_{i}({\bm k}) + \imath
{\bm {\hat{\sigma}}} \cdot {\bm D}_{i}({\bm k})$, equations (\ref{8}), (\ref{9})
can be rewritten as follows :   
\begin{eqnarray}
&& \langle \chi_{1,S_z} |{\cal{J}}^{+}(K\vec{e}_z,m_d,m_d)|\chi_{1,S_z}\rangle =
\sqrt{2}m_d {(\vec {e}_{S_z})}_{j'}^{*} {(\vec {e}_{S_z})}_{j}
\int\nolimits
\left[ \frac{\omega (k) \xi'}{\omega (k') \xi} \right] ^{1/2}
F_{i'j'}({\bm k}')  F_{ij}({\bm k})  \cdot
\nonumber\\
&&\frac{1}{2}Tr \left\{ \left[ C_{i'}({\bm k}') - 
\imath {\bm {\hat{\sigma}}} \cdot {\bm D}_{i'}({\bm k}') \right] 
\left[ A^+ + \imath {\bm {\hat{\sigma}}} \cdot {\bm B^+} \right] 
\left[ C_{i}({\bm k}) + 
\imath {\bm {\hat{\sigma}}} \cdot {\bm D}_{i}({\bm k}) \right] \right\} 
d{\bm k}'  
\label{59'} 
\end{eqnarray} 
\begin{eqnarray}
&&\langle \chi_{1,S_z'} |{\cal{J}}_{x}(K\vec{e}_z,m_d,m_d)|\chi_{1,S_z}\rangle =
{(\vec {e}_{S_z'})}_{j'}^{*} {(\vec {e}_{S_z})}_{j}
\int\nolimits
\left[ \frac{\omega (k)}{\omega (k')} \right] ^{1/2}
F_{i'j'}({\bm k}')  F_{ij}({\bm k})  \cdot
\nonumber\\
&&\frac{1}{2}Tr \left\{ \left[C_{i'}({\bm k}') - 
\imath {\bm {\hat{\sigma}}} \cdot {\bm D}_{i'}({\bm k}') \right] 
\left[ A_x + \imath {\bm {\hat{\sigma}}} \cdot {\bm B_x} \right] 
\left[ C_{i}({\bm k}) + 
\imath {\bm {\hat{\sigma}}} \cdot {\bm D}_{i}({\bm k}) \right] \right\} 
\frac{d{\bm k}'}{\xi}  
\label{59} 
\end{eqnarray} 
where 
\begin{eqnarray}
A^+ + \imath {\bm {\hat{\sigma}}} \cdot {\bm B^+} =
\frac{am(f_e^{is} - f_m^{is}) 
\left[ am + \imath b({\hat{\sigma}} k)_{\bot} \right] }
{a^2m^2+ b^2k_{\bot}^2} + f_m^{is}   
\label{62'}
\end{eqnarray}
and
\begin{eqnarray}
A_x + \imath {\bm {\hat{\sigma}}} \cdot {\bm B_x} =
\frac{4mk_x(f_e^{is}-f_m^{is})[am + \imath b({\hat{\sigma}} k)_{\bot}]}
{a^2m^2 + b^2k_{\bot}^2} +  f_m^{is} [ak_x + \imath b(m{\hat{\sigma}}_y
+ k_y {\hat{\sigma}}_z)] . 
\label{62}
\end{eqnarray}
It is straightforward to see that $A_x$ is proportional to the
quantity $a$, while ${\bm B^+}$ and ${\bm B_x}$ are proportional to $b$,
defined in Eq. (\ref{4}). All the quantities $A^+, {\bm B^+}, A_x, {\bm B_x}, 
C_{i}({\bm k}), {\bm D}_{i}({\bm k})$ are real. 

By an explicit calculation of the traces in Eqs. (\ref{59'}),
(\ref{59}) one has  
\begin{eqnarray} 
&& \langle \chi_{1,S_z} |{\cal{J}}^{+}(K\vec{e}_z,m_d,m_d)|\chi_{1,S_z}\rangle = 
\sqrt{2}m_d {(\vec {e}_{S_z})}_{j'}^{*} {(\vec {e}_{S_z})}_{j} 
\int\nolimits
\left[ \frac{\omega (k) \xi'}{\omega (k') \xi} \right] ^{1/2} 
F_{i'j'}({\bm k}')  F_{ij}({\bm k}) \cdot 
\nonumber\\ 
&&\left\{ A^+ \left[ C_{i'}({\bm k}') C_{i}({\bm k}) + 
{\bm D}_{i'}({\bm k}') \cdot {\bm D}_{i}({\bm k}) \right] - \right.
\nonumber \\
&& \left. {\bm B^+} \cdot \left[  C_{i'}({\bm k}') {\bm D}_{i}({\bm k}) -
{\bm D}_{i'}({\bm k}') C_{i}({\bm k}) -
{\bm D}_{i'}({\bm k}') \wedge {\bm D}_{i}({\bm k}) 
\right] \right\}
d{\bm k}'  
\label{63'} 
\end{eqnarray} 
\begin{eqnarray}
&&\langle \chi_{1,S_z'} |{\cal{J}}_{x}(K\vec{e}_z,m_d,m_d)|\chi_{1,S_z}\rangle =
{(\vec {e}_{S_z'})}_{j'}^{*} {(\vec {e}_{S_z})}_{j}
\int\nolimits
\left[ \frac{\omega (k)}{\omega (k')} \right] ^{1/2}
F_{i'j'}({\bm k}')  F_{ij}({\bm k})  \cdot
\nonumber\\
&&\left\{ A_x \left[ C_{i'}({\bm k}') C_{i}({\bm k}) + 
{\bm D}_{i'}({\bm k}') \cdot {\bm D}_{i}({\bm k}) \right] - \right.
\nonumber \\
&&\left. {\bm B_x} \cdot \left[  C_{i'}({\bm k}') {\bm D}_{i}({\bm k}) -
{\bm D}_{i'}({\bm k}') C_{i}({\bm k}) -
{\bm D}_{i'}({\bm k}') \wedge {\bm D}_{i}({\bm k})
\right] \right\}
\frac{d{\bm k}'}{\xi}  
\label{63} 
\end{eqnarray} 
It has to be noted that the integrals in Eqs. (\ref{63'}, \ref{63}) are real.
Therefore, since the matrix elements ${\cal{J}}^{+}_{S_z,S_z}$ and
${\cal{J}}^{x}_{S_z',S_z}$ are real (see the end of Sect. 2), only the real part
of ${(\vec {e}_{S_z'})}_{j'}^{*}  {(\vec {e}_{S_z})}_{j}$ can contribute to these
matrix elements.

\section{Numerical results for the deuteron form factors} 
\label{S5}
\subsection{Deuteron magnetic and quadrupole moments} 
\label{S50}
The direct evaluation of magnetic and quadrupole moments through the limits of
Eqs. (\ref{99'},\ref{101}) implies very delicate numerical problems
and then a careful analitical reduction of these equation is needed.
For the sake of completeness we report in Appendix D the explicit expressions
that have actually been used. Magnetic and quadrupole moments have already been
calculated in Ref. \cite{LPS3} for a variety of $N-N$ interactions. In this
paper we recall our main results, which are summarized in Table I. In the table
the values of the magnetic and quadrupole moments calculated with many $N-N$
interactions, already shown in Ref. \cite{LPS3}, are reported together with the
values obtained using the local $Nijmegen2$ interaction, which was not
considered in Ref. \cite{LPS3}.

The standard non-relativistic
results obtained with a one-body current crucially depend on the asymptotic
normalization ratio $\eta = A_D/A_S$ of $D$ and $S$ wave functions and on the
$D-$state percentage in the deuteron, $P_D$, but one cannot obtain at the same
time the experimental values for both $\mu_d$ and $Q_d$. Using the free current
within the FFHD in the $q^+=0$ reference frame, the relativistic correction (RC)
turned out to be very small for $Q_d$, while for $\mu_d$ it could explain only
part of the disagreement with the experimental value \cite{CCKP}. On the
contrary, in our Poincar\'e covariant calculation \cite{LPS3} the RC's bring
both $\mu_d$ and $Q_d$ closer to the experimental values, except for the 
charge-dependent Bonn interaction \cite{CDB}. We wish to stress that our current
operator and the one used in Ref. \cite{CCKP} are different, since both of them
are obtained from the free one, but in different reference frames, related by an
interaction dependent rotation. As was already observed for the nonrelativistic
calculations of $Q_d$ \cite{Rosa,Klar}, we have shown in Ref. \cite{LPS3} that a
remarkable linear behaviour against the asymptotic normalization ratio, $\eta$,
holds for both the deuteron moments calculated within our approach (the values
of $\mu_d$ and $Q_d$ corresponding to the $Nijmegen2$ interaction obey
precisely the same trend as the other interactions). The values of $\mu_d$ and
$Q_d$, suggested by this linear behaviour in correspondence of the experimental
value of $\eta$ ($\eta^{exp} =  0.0256(4)$ \cite{eta}) differ from the
experimental ones ($\mu_d = 0.857406(1)$ \cite{Lind} and $Q_d = 0.2859(3)$
\cite{Rosa}) only by $0.5\%$ and $2\%$, respectively, i.e. much less than for
the non-relativistic results. The RC to $\mu_d$ is rather large and the total
result becomes slightly greater than $\mu_d^{exp}$, while the nonrelativistic
one is smaller. This shows that, within our framework, even the sign of explicit
contributions of two-body currents is different from the one needed in the
non-relativistic case. In conclusion, it appears that, within our approach, the
total contribution of two-body currents (from meson-exchange, Z-graphs, etc.)
and isobar configurations has to be relatively small at  $Q^2 = 0$.

\subsection{Deuteron form factors and $N-N$ interactions} 
\label{S51}

Let us first compare in Figs. 1 and 2 our relativistic results for $A(Q^2)$,
$B(Q^2)$ and $T_{20}(Q^2)$, obtained using the $RSC$ interaction \cite{RSC} and
the Gari-Kr\"{u}mpelmann nucleon form factors \cite{Gari}, with the corresponding
nonrelativistic results. Following Lomon \cite {Lomon}, the latter ones have been
obtained by using the exact relativistic relations between the deuteron form
factors and the current matrix elements, within the Breit reference frame where
the momentum transfer is directed along the $z$ axis \cite{Gourdin}, but
with nonrelativistic expressions for the matrix elements evaluated in
impulse approximation \cite {Lomon}. 

In order to have a closer insight to the form
factor behaviour, in addition to the usual plots for $A(Q^2)$ and
$B(Q^2)$ in a logarithmic scale, shown in Fig. 1, we report in Fig. 2(a) the
quantity  $A(Q^2)$ divided by the factor $(G_D^2 \cdot F)$, with  $G_D = (1 +
Q^2/0.71)^{-2}$ and $F = (1 + Q^2/0.1)^{-2.5}$, in a linear scale,  and in Fig.
2(b) the quantity $\Gamma_M (Q^2) = [G_M (Q^2) m_p / (\mu_d m_d)]^2$ divided by
the factor $(G_D^2 \cdot F_1)$, with $F_1 = (1 + Q^2/0.1)^{-3}$. As it is clear
from Figs. 1 and 2, the differences between relativistic and nonrelativistic
results are a few percent for $Q^2 \leq 0.1 (GeV/c)^2$, while become large as
$Q^2$ increases. For $A(Q^2)$ the differences are larger than $20 \%$ already at
$Q^2 \geq 0.2 (GeV/c)^2$ and are of orders of magnitude for $Q^2 \geq 2
(GeV/c)^2$. For $B(Q^2)$ the relativistic and nonrelativistic results differ by
$50 - 100 \%$ for  $Q^2 \geq 0.3 (GeV/c)^2$, while for $T_{20}(Q^2)$ they
considerably differ for  $Q^2 \ge 0.5 (GeV/c)^2$. In Figs. 1, 2 we have also
reported by dashed lines the results obtained by keeping fixed the argument of
the nucleon form factors in Eqs.  (\ref{8},\ref{9}). The effects of
factorization become large for $A(Q^2)$ and $B(Q^2)$ at $Q^2 \ge 1 (GeV/c)^2$,
while for $T_{20}(Q^2)$ already at $Q^2 \ge 0.5 (GeV/c)^2$. 
From Fig. 1
it appears that the nonrelativistic approach is able to give an overall
description of the data for $A(Q^2)$, $B(Q^2)$, and $T_{20}(Q^2)$. However, this
description is not accurate, even at very low values of the momentum transfer, as
one can see in Fig. 2 (a) and, furthermore, it strongly depends on the $N-N$
interaction and the nucleon form factor model. For instance,
using the $CD-Bonn$ interaction \cite{CDB} and the nucleon form factors by
Hoehler et al. \cite{Hoehler}, for $A(Q^2)$ and $T_{20}(Q^2)$
the agreement is completely lost at $Q^2 \ge 0.4 (GeV/c)^2$.

A comparison of our results with the deuteron form factors obtained by using the
same $N-N$ interactions and the same nucleon form factors, but within different
relativistic approaches, for instance within the front-form calculation of Ref. 
\cite{CCKP}, can also be interesting. Using the Paris interaction \cite{Par}
and the form factors of Ref. \cite{Hoehler}, large differences have been found
for $A(Q^2)$ at $Q^2 \geq 2 (GeV/c)^2$, which become of orders of magnitude at
$Q^2 = 6 (GeV/c)^2$ (see Ref. \cite{LPS2}). For $B(Q^2)$ we found a minimum
around $Q^2 = 1.8 (GeV/c)^2$ instead of $Q^2 = 1.6 (GeV/c)^2$ as in  Ref.
\cite{CCKP}, and for $T_{20}(Q^2)$ a zero at $Q^2 = 1.4 (GeV/c)^2$ instead of
$1.2 (GeV/c)^2$.

The results obtained within our approach with different $N - N$ interactions are
analyzed in Figs. 3 and 4, using the nucleon form factor model by Hoehler et al.
\cite{Hoehler}. We consider the old RSC interaction \cite{RSC} and recent
realistic interactions, able to describe the two-body data with a reduced $\chi
^2 \approx 1$. In particular we study the $AV18$ interaction by the Argonne
group  \cite{AV18}, some interactions by the Nijmegen group ($Nijmegen1$,
$Nijmegen2$, $Nijmegen 93$, $Reid93$) \cite{Nij}, and the  charge-dependent
$CD-Bonn$ interaction by the Bonn group \cite{CDB}. The results for the
$Reid93$ interaction are essentially equal to the results of the 
$AV18$ interaction and are not reported in the figures.

The effects of different
interactions are large for  $A(Q^2)$ at $Q^2 \ge 1 (GeV/c)^2$, while for
$B(Q^2)$ and $T_{20}(Q^2)$ already at $Q^2 \ge 0.5 (GeV/c)^2$. It can be noted
that the $CD-Bonn$ interaction, which is characterized by a larger non-locality,
yields larger differences with respect to the other interactions. At low values
of $Q^2$ ($Q^2 < 0.4 (GeV/c)^2$), where the nucleon form factors are better
known, a simultaneous description of the experimental data for $A(Q^2)$,
$B(Q^2)$ and  $T_{20}(Q^2)$ is achieved. The dependence on the nucleon-nucleon
interaction in this region is minor, although not negligible (see, in
particular, Fig. 4 (b)).

For the mentioned interactions and using the
Gari-Kr\"{u}mpelmann nucleon form factors \cite{Gari}, we report in Fig. 5 (a)
the value of $Q^2$ corresponding to the minimum of $B(Q^2)$ and in Fig. 5 (b)
the value of $Q^2$ corresponding to the second zero of $T_{20}(Q^2)$ 
against the nonrelativistic S-state kinetic energy, $T_S$, in order to find a
correlation between different effects of the $N - N$ interactions. For both
quantities a distinct linear behaviour is clear: a lower value of $T_S$ yields a
minimum for $B(Q^2)$ and a zero for $T_{20}(Q^2)$ at a larger momentum transfer.
Analogous results can be obtained with different nucleon form factors, as the
ones of Ref. \cite{Hoehler}.  From Figs. 3(a), 4(a) it is clear that for  $Q^2
\ge 1 (GeV/c)^2$ a similar correlation holds for $A(Q^2)$, i.e. a lower value of
$T_S$ yields a lower value of $A(Q^2)$. It has also to be noted that the $AV18$
and $Reid93$ interactions, which give essentially the same results for $A(Q^2)$,
$B(Q^2)$ and $T_{20}(Q^2)$, have the same S-state kinetic energy.

Let us note that recent measurements of the $S-D$ mixing parameter,
$\epsilon_1$, point to a stronger tensor force than the one
exhibited by the interaction models we have analyzed \cite{Tornow}.
In turn, a stronger tensor force is favoured by a high degree of locality,
which yields significantly larger kinetic energies and, in particular, larger
values of $T_S$ \cite{Polls}. Then,
 by an extrapolation of the linear relations found above, one can argue that a
$N-N$ interaction able to reproduce these recent measurements of $\epsilon_1$
could yield, on one hand, agreement between experimental and
theoretical values for $T_{20}(Q^2)$ and, on the other one, a
minimum for $B(Q^2)$ slightly lower than the value indicated by the available
experimental data (around $Q^2 = 1.6 (GeV/c)^2$ instead of $Q^2 =
1.8 (GeV/c)^2$). Therefore, if new, more precise experimental data for $B(Q^2)$
will show such a lower value for the position of the minimum, both $B(Q^2)$ and
$T_{20}(Q^2)$ could be reproduced by a novel $N-N$ interaction, without a
relevant role for explicit two-body currents.

\subsection{Deuteron form factors and nucleon electromagnetic form factors}
\label{S52}
In order to investigate the effects of the nucleon form factors on the deuteron
form factors, we have displayed in Fig. 6 our results obtained with the 
$Nijmegen2$
nucleon-nucleon interaction and corresponding to the nucleon form factor models
of Refs. \cite{Gari}, \cite{Hoehler}, and \cite{Mergell}. For $A(Q^2)$ the
differences between different models are very large at $Q^2 \ge 0.5 (GeV/c)^2$,
increase as $Q^2$ increases, and can be related to the sizeably different
behaviour of $G_E^n(Q^2)$ for the various models. The
influence of the  nucleon form factor models is less marked in $B(Q^2)$, while,
as already known \cite{Schia}, the tensor polarization is essentially
independent of the nucleon form factors. 

Therefore, the linear behaviour of the locations of the minimum of $B(Q^2)$
and the second zero of $T_{20}(Q^2)$ vs. $T_S$ is substantially independent
of the form factor models, as well as the conjecture at the end of the previous
paragraph. As far as $A(Q^2)$ is concerned, one could try to exploit the strong
dependence of $A(Q^2)$ on $G_E^n(Q^2)$ to gain information on $G_E^n(Q^2)$
 by a fit of the $A(Q^2)$ experimental data, following a procedure analogous to
the one used, in a nonrelativistic context, by Platchkov et al. \cite{Saclay}.
Obviously the results of this fit will be different for different interactions.
Another possibility to be studied in our covariant framework is obviously the
role of isobar configurations in the deuteron state (see, e.g.\cite{Despla}) and
of explicit two-body contributions in the e.m. current (see, e.g. \cite{Hummel}).
As already noted \cite{LPS},\cite{LPS1}, these contributions have to be
Poincar\'e covariant, and to satisfy Hermiticity and current conservation by
themselves. We intend to perform such a fit and to study these contributions
elsewhere.

\section{Conclusions} 
\label{S6}
In this paper the deuteron form factors $A(Q^2)$ and $B(Q^2)$, and the tensor
polarization $T_{20}(Q^2)$ have been evaluated in the framework of front-form
Hamiltonian dynamics, using a Poincar\'e covariant current operator, without
any ambiguity. The current is built up from the free one in the Breit reference
frame where $\vec{q}$ is along the $z$ axis and fulfills parity and time reversal
covariance, as well as Hermiticity and current conservation.

Large differences have been found between the
results of calculations performed within a nonrelativistic framework and within
our Poincar\'e covariant approach. These differences become huge at high momentum
transfer, as expected, but are relevant for accurate calculations even in the
limit of zero momentum transfer, as is clear from our results for the deuteron
magnetic and quadrupole moments \cite{LPS3}. Large differences have also been 
found with respect to a front-form approach which ensures Poincar\'e covariance
by different definitions for different matrix elements of the current
operator (\cite{CCKP}).  
Our current operator, which was already shown to be able to describe the deuteron
magnetic and quadrupole moments, is also able to simultaneously reproduce the
three deuteron form factors at low momentum transfer, where the nucleon form
factor are better known and the effects of different interactions are minor.

 The effects on the deuteron form
factors of different nucleon-nucleon interactions and different nucleon form
factor models have been studied. The different nucleon form
factor models strongly affect $A(Q^2)$, while the different interactions have
large effects on $A, B$ and $T_{20}$. These effects are linked to the $S$-state
kinetic energy in the deuteron, which, in turn, is related to the degree of
non-locality of the interactions and to the strenght of the tensor force. A
novel $N-N$ interaction with a strong tensor force, able to reproduce the recent
measurements of $\epsilon_1$, would be helpful to describe the deuteron form
factors and, in particular, to offer a solid ground for the study of the neutron
charge form factor from the analysis of $A(Q^2)$.
We stress the relevance of a well defined
relativistic approach to gain reliable information on the nucleon-nucleon
interaction and the nucleon form factors.

\section{Acknowledgements} 
\label{S7}

The authors wish to thank A. Kievsky for kindly providing the
deuteron wavefunctions for RSC, and Av18 interactions and R.
Machleidt for the CD-Bonn wavefunction. The wave functions for the Nijmegen
interactions have been downloaded from: http://nn-online.sci.kun.nl \,\,\,\,.
This work was partially supported by Ministero della Ricerca
Scientifica e Tecnologica. 

\begin{center}
\Large{\bf Appendix}
\end{center}
In this Appendix we report for the sake of completeness the explicit
expressions of some useful quantities needed for the calculation of the deuteron
em form factors.
\appendix 
\section{Polarization vectors} 
\label{S8}
The deuteron polarization four-vectors, $e_{S_z}$, in any reference
frame can be obtained by a proper boost from the polarization vectors in the
deuteron rest frame, $e_{S_z}(rf) \equiv (e_{rf}^0 = 0, {\vec e_{S_z}})$, 
with

\begin{eqnarray} 
{\vec e}_{+1} = - \frac{1}{\sqrt{2}} \, (1, \, \imath, \, 0), 
\quad {\vec e}_{-1} = \frac{1}{\sqrt{2}} \, (1, \, - \imath, \, 0),\quad 
{\vec e}_{0} = (0, \, 0, \, 1).
\label{35}
\end{eqnarray}
In our Breit frame, where ${\vec P}_{\bot} = {\vec P}_{\bot}' = 0$, the
transverse deuteron polarization vectors, in both the initial and final states,
read as follows: 
\begin{eqnarray} 
e_{\pm 1} = e_{\pm 1}' = 
\mp \frac{1}{\sqrt{2}} \, ( 0, \, 1, \, \pm \imath, \, 0 ) , 
\label{36}
\end{eqnarray} 
while the longitudinal polarization vector in the initial state is
\begin{eqnarray} 
e_{0} = \frac{1}{m_d} \, ( - K, \, 0, \, 0, \, \sqrt{m_d^2 + K^2} )
\label{37}
\end{eqnarray}
and in the final state is
\begin{eqnarray} 
e_{0}^{'} = \frac{1}{m_d} \, ( K, \, 0, \, 0, \, \sqrt{m_d^2 + K^2} ).
\label{40}
\end{eqnarray}

\section{Front-form Dirac spinors and matrix elements of $\gamma$ matrices} 
\label{S9}
Adopting the following representation for the $\gamma$ matrices
\begin{eqnarray} 
\gamma ^0 = \left|\left| \begin{array}{cc} 0 & 1 \\ 1 & 0 \end{array}
\right|\right|,  
\quad  
\gamma ^5 = \left|\left| \begin{array}{cc} 1 & 0 \\ 0 & -1 \end{array}
\right|\right|, 
\quad  
\gamma ^i = \left|\left| \begin{array}{cc} 0 & -\sigma_i \\ \sigma_i & 0 
\end{array} \right|\right| , 
\label{135}
\end{eqnarray}
with $i=1,2,3$ and $\sigma_i$ the Pauli matrices, the front-form Dirac spinor
$w(p,\sigma)$ can be written as
\begin{eqnarray} 
w(p,\sigma) = \sqrt{m} \left|\left| \begin{array}{c} \beta (g) \chi (\sigma)  \\
\left[ \beta (g)^{-1} \right] ^{\dagger} \chi(\sigma)  \end{array}
\right|\right|,   
\label{136}
\end{eqnarray}
where $\chi(\sigma)$ is the ordinary spin $1/2$ spinor describing the state with
spin projection on the $z$ axis equal to $\sigma$ and the matrix $\beta (g)$ has
the components
\begin{eqnarray} 
\beta_{11} = \beta_{22}^{-1} = 2^{1/4} (g^+)^{1/2}, 
\quad \beta_{12} = 0,
\quad 
\beta_{21} = (g_x + g_y) \beta_{22} ,
\label{137}
\end{eqnarray}
with $g = p/m$.

One can immediately obtain
\begin{equation}
\bar{w}(p',\sigma') w(p,\sigma)  = \frac {1} {\sqrt{p^+ p'^+}}
\langle \sigma'| \left [ m (p^+ + p'^+) - 
\imath {\hat{\sigma}}_x (p^+ p'_y - p'^+ p_y) + 
\imath {\hat{\sigma}}_y (p^+ p'_x - p'^+ p_x) \right]| \sigma \rangle ,  
\label{118}
\end{equation}
with normalization
\begin{equation}
\bar{w}(p,\sigma') w(p,\sigma)  = \frac {1} {p^+}
\langle \sigma'| m 2 p^+ | \sigma \rangle = 2 m \delta_{\sigma \sigma '} .  
\label{118'}
\end{equation}
The matrix elements of the $\gamma$ matrices, needed for the calculation of
the deuteron form factors, are
\begin{eqnarray} 
&\bar{w}(p',\sigma') \gamma ^+ w(p,\sigma) = 2 \sqrt{p^+ p'^+} \delta_{\sigma
\sigma '} 
\label{129} 
\\
\nonumber\\  
&\bar{w}(p',\sigma') \gamma _x w(p,\sigma) = \frac {1} {\sqrt{p^+ p'^+}}
\langle \sigma'| \left [ \imath m q^+ {\hat{\sigma}}_y + p^+ p'_x + p'^+ p_x 
+ \imath {\hat{\sigma}}_z (p'^+ p_y - p^+ p'_y) \right]| \sigma \rangle. 
\label{138}
\end{eqnarray}
In our special Breit frame Eqs. (\ref{118}) and (\ref{138}) become :
\begin{eqnarray} 
\bar{w}(p',\sigma') w(p,\sigma) = \frac {1} {\sqrt{p^+ p'^+}}
\langle \sigma'| \left [ m (p^+ + p'^+) + \imath q^+
({\hat{\sigma}}_x k_y - {\hat{\sigma}}_y k_x ) \right]| \sigma \rangle 
\label{120} 
\\  
\bar{w}(p',\sigma') \gamma _x w(p,\sigma) = \frac {1} {\sqrt{p^+ p'^+}} \langle
\sigma'| \left [ \imath m q^+ {\hat{\sigma}}_y + (p^+ + p'^+) k_x  + \imath
{\hat{\sigma}}_z k_y q^+  \right]| \sigma \rangle.  
\label{139}
\end{eqnarray}

\section{Generalized Melosh matrices for the deuteron wave function}
\label{S12}

The generalized Melosh matrices for the deuteron wave function have been
defined in Sect. 4 as the matrices 
\begin{eqnarray} 
C_{i}({\bm k}) + \imath {\bm {\hat{\sigma}}} \cdot {\bm D}_{i}({\bm k}) =
v({\bm k},\vec s_1)^{-1}{\hat{\sigma}}_{i}{\hat{\sigma}}_y [v(-{\bm k},\vec
s_2)]^*, \quad i = 1, 2, 3 .
\label{149}
\end{eqnarray}
From the expression (\ref{27}) for the matrix $v({\bm k},\vec s)$ one obtains
\begin{eqnarray} 
&&C_{i}({\bm k}) = {\cal N} \left[ \delta_{2i} m + 
\frac {k_yk_i} {m + \omega (k) } \right]  
\label{169} 
\\
&&[{\bm D}_{i}({\bm k})]_x = {\cal N} \left[ - \delta_{3i} m + 
\frac { k_z k_i - \delta_{3i} k^2 } {m + \omega (k)} \right]   
\label{179}
\\
&&[{\bm D}_{i}({\bm k})]_y = {\cal N} ({\bm e}_z \wedge {\bm k})_i 
\label{189}
\\
&&[{\bm D}_{i}({\bm k})]_z = {\cal N} \left[ \delta_{1i} m + 
\frac { k_x k_i} {m + \omega (k)} \right] 
\label{199}
\\
\end{eqnarray}
where
\begin{eqnarray}
{\cal N} = \frac{1}{M_0 \sqrt{\xi(1-\xi)}} = \frac{1}{ \sqrt{m^2 + k_{\bot}^2}}
\label{159}
\end{eqnarray}

\section{Deuteron magnetic and quadrupole moments}
\label{S10}
In this Appendix we illustrate the main steps for the calculations of
the deuteron magnetic and quadrupole moments from Eqs. (\ref{99'},\ref{101}).
To this end, expansions in $\kappa = \sqrt{\tau} = Q/(2m_d)$ of the quantities
$a$ and $b$ (Eq. (\ref{4})) up to the first order
\begin{eqnarray}
a = 2 , 
\quad  b = 2 \frac{\kappa}{\xi}    
\label{19'}
\end{eqnarray}
and of the quantities $\xi$ and $k_z$ (Eq. (\ref{19})) up to the second order
\begin{eqnarray}
\xi = \xi' - 2 \kappa (1 - \xi') - 2 \kappa^2 (1 - \xi'),
\quad  k_z = k_z' - \frac{\kappa \omega(k')}{\xi'} +
\omega(k')\frac{\kappa^2}{2 \xi^{'2}}(4\xi' - 3)
\label{20'}
\end{eqnarray}
will be needed, since the intrinsic moment in the final state, ${\vec k}'$, is
the integration variable in the integrals for the calculation of the current
matrix elements.

\subsection{Magnetic moment} 
The deuteron magnetic moment is given by Eq. (\ref{99'})
\begin{equation}
\mu _d = \frac{m_p}{(\sqrt{2} m_d)} \lim_{Q \rightarrow 0}   \frac{1}{ Q}
[{\cal{J}}^{x}_{1,0} - {\cal{J}}^{x}_{0,1}] \: ,
\label{99''}
\end{equation}
where the matrix elements ${\cal{J}}^{x}_{1,0}$ and ${\cal{J}}^{x}_{0,1}$ can
be obtained by Eq. (\ref{63}). Let us preliminarly note that 
${\cal{J}}^{x}_{1,0}$ and ${\cal{J}}^{x}_{0,1}$ have the same expression, but for
the exchange of the role of initial and final variables in $F_{i,j}({\bm k})$,
$F_{i',j'}({\bm k}')$, and in the quantity between curly brackets in Eq.
(\ref{63}) (we recall that only the real part of ${(\vec {e}_{+1})}_{j}$ gives a
non vanishing contribution to the matrix elements).
In order to obtain the magnetic moment, one can expand 
$[{\cal{J}}^{x}_{1,0} - {\cal{J}}^{x}_{0,1}]$ as a
function of $\kappa$, and consider only the terms which are linear in
$\kappa$ (indeed, because of Eqs. (\ref{19}) and (\ref{119}), the current matrix
elements are functions of $\kappa$). 
As a first step, by using Eq. (\ref{19'}) and Eq.
(\ref{20'}), we expand the quantity between
curly brackets in Eq. (\ref{63}) at the first order in $\kappa$. We obtain a term
independent of $\kappa$ and a term linear in $\kappa$, which is
identical, but with opposite signs, for the two matrix elements
${\cal{J}}^{x}_{1,0}$ and ${\cal{J}}^{x}_{0,1}$. It is clear that in
correspondence to the latter term one can evaluate the radial wave functions in
Eq. (\ref{63}) with the same argument ${\bm k}$. After an integration over the
polar angle $\phi$  [${\bm k}\equiv (k,\theta,\phi)$] one has:   
\begin{eqnarray}
&&\mu_d = - \lim_{Q \rightarrow 0} \frac{m_p [ {\cal{F}} - {\cal{F}'} ]}
{ 2 Q m_d} \; + \nonumber\\ 
&&\frac{\pi  m m_p}{m_d^2}
\int_{0}^{\infty} d(k_{\bot}^2)
\int\nolimits_{-\infty}^{\infty}\frac{dk_z}{\xi^2} \,
f_m^{is} \chi_0(k)
\left[\chi_0(k) \left(1 + \frac{k_{\bot}^2}{2m(\omega(k)+m)} \right) +
3\varphi_2(k)\frac{k^2+k_z^2}{2\sqrt{2}k^2} \right] + \nonumber\\
&&\frac{m_p \pi}{2 m m_d^2}\int_{0}^{\infty} d(k_{\bot}^2) \, k_{\bot}^2
\int\nolimits_{-\infty}^{\infty}\frac{dk_z} {\xi^2} \,
\chi_0(k) \left[ \chi_0(k) + \frac{3}{\sqrt{2}}\varphi_2(k) \right] \left[
f_m^{is} - f_e^{is} \frac{\omega(k)}{\omega(k) + m} \right] 
\label{91}
\end{eqnarray}
where the first term and the last two terms correspond to the zero and first
order terms in the expansion of the curly bracket of Eq.
(\ref{63}), respectively. In Eq. (\ref{91}), $\cal{F}$ is given by the following
expression \begin{eqnarray} &&{\cal{F}} = 3 \pi \int_{0}^{\infty} d(k_{\bot}^2)
\, k_{\bot}^2 \int_{-\infty}^{\infty} \frac{dk_z'}{\xi}
\left[ \frac{\omega(k)}{\omega(k')} \right]^{1/2} f_e^{is} \cdot \nonumber\\
&&\left [\chi_0(k)\varphi_2(k')\frac{k_z'}{\sqrt{2}k^{'2}}+
\chi_0(k')\varphi_2(k)\frac{k_z}{\sqrt{2}k^{2}} + 
\frac{3k_z(k_{\bot}^2+k_zk_z')}{2k^2k^{'2}}
\varphi_2(k)\varphi_2(k')\right]
\label{81}
\end{eqnarray}
and, according to the observation at the beginning of this subsection, 
$\cal{F}'$ has the same expression, but for the exchange of ${\bm k}$ and 
${\bm k}'$ in the quantity between square brackets. 

The limit in Eq. (\ref{91}) can be easily handled and one obtains the final
result 
\begin{eqnarray}
&&\mu_d = 8 \pi \frac {m m_p}{m_d^2}
\int_{0}^{\infty} k^2 d k
\int_{0}^{1} d(cos \theta) \frac{[(\omega(k))^2 + k_z^2]}{(m^2 + k_{\bot}^2)^2}
\; \cdot \nonumber\\
&&\left\{ \frac{9 \omega(k)}{4 m} [\varphi_2(k)]^2 (1 - cos^2 \theta) +
f_m^{is} \chi_0(k) 
\left[ \chi_0(k) \left( 2 + \frac{k_{\bot}^2}{m(\omega(k) + m)} \right) +
3 \varphi_2(k)\frac{(1 + cos^2 \theta)}{\sqrt{2}} \right] \right.+ \nonumber\\ 
&& \left. \chi_0(k) \frac {k_{\bot}^2}{m^2} 
\left[ \chi_0(k) + \frac{3}{\sqrt{2}}\varphi_2(k) \right] \left[ f_m^{is} -
\frac {\omega(k)} {\omega(k) + m} \right]  \right\} \; .
\label{82} 
\end{eqnarray}

In Eqs. (\ref{91},\ref{82}) the nucleon form factors $f_e^{is}$ and $f_m^{is}$ 
have to be evaluated in the limit $Q \rightarrow 0$, i.e. $f_e^{is}(0) = 1$,
$f_m^{is}(0) = 0.8797$.

The nonrelativistic result for $\mu_d$ can be immediately recovered from Eq. 
(\ref{82}) in the limit $m \rightarrow \infty$.

\subsection{Quadrupole moment}
The quadrupole form factor (see Eq. (\ref{98'''})) is given by
\begin{eqnarray}
 G_Q = \frac{\sqrt{2} m_d}{Q^2}  
\frac {[ {\cal{J}}^{+}_{0,0} - {\cal{J}}^{+}_{1,1} ]}{\sqrt{1 + \tau}} \; . 
\label{111}
\end{eqnarray}
The proper combination $( {\cal{J}}^{+}_{0,0} - {\cal{J}}^{+}_{1,1} )$ of the
matrix elements of ${\cal{J}}^{+}$ can be directly calculated from 
Eq. (\ref{63'}) by using Eqs. (\ref{60'}), (\ref{62'}) and the explicit
expressions for the the quantities $C_{i}({\bm k}), {\bm D}_{i}({\bm k})$ 
given in Appendix C.
One obtains
\begin{eqnarray}
{\cal{J}}^{+}_{0,0} - {\cal{J}}^{+}_{1,1} = 
m_d \sqrt{2} \int\nolimits \left[ \frac{\omega(k)\xi'}{\omega (k')\xi} \right]
^{1/2} \left[ (f_e^{is} - f_m^{is}) bk_{\bot} 
\frac {(b k_{\bot} E - am H)}{a^2m^2 + b^2k_{\bot}^2} - f_e^{is} E \right]
d{\vec k}'
\label{D1}
\end{eqnarray}
where
\begin{eqnarray}
&&E = \frac{1}{2}\chi_0(k)\chi_0(k')
[1-cos(\varphi-\varphi')] + \nonumber\\
&&\frac{3\varphi_2(k)\chi_0(k')}{\sqrt{2}k^{2}}
\left[ (\frac{1}{2}k_{\bot}^2-k_z^2)cos(\varphi-\varphi')+\frac{3}{2}k_z
k_{\bot}sin(\varphi-\varphi') \right] + \nonumber\\
&&\frac{3\varphi_2(k')\chi_0(k)}{\sqrt{2}k^{'2}}
\left[ (\frac{1}{2}k_{\bot}^2-k_z^{'2})cos(\varphi-\varphi')-\frac{3}{2}k_z'
k_{\bot}sin(\varphi-\varphi') \right] + \nonumber\\
&&\frac{9\varphi_2(k)\varphi_2(k')}{2k^2k^{'2}}
(\frac{1}{2}k_{\bot}^2-k_zk_z')[(k_{\bot}^2+k_zk_z')
cos(\varphi-\varphi') +
(k_z-k_z')k_{\bot}sin(\varphi-\varphi')]
\label{D2}
\end{eqnarray}
and
\begin{eqnarray}
&&H = \frac{1}{2}\chi_0(k)\chi_0(k')
sin(\varphi-\varphi') + \nonumber\\
&&\frac{3\varphi_2(k)\chi_0(k')}{\sqrt{2}k^{2}}
\left[ (k_z^2 - \frac{1}{2}k_{\bot}^2) sin(\varphi-\varphi') + 
\frac{3}{2} k_zk_{\bot}cos(\varphi-\varphi') \right] + \nonumber\\
&&\frac{3\varphi_2(k')\chi_0(k)}{\sqrt{2}k^{'2}})
\left[ (k_z^{'2} - \frac{1}{2}k_{\bot}^2) sin(\varphi-\varphi')
- \frac{3}{2} k_z'k_{\bot} cos(\varphi-\varphi') \right] + \nonumber\\
&&\frac{9\varphi_2(k)\varphi_2(k')}{2k^2k^{'2}}
(k_zk_z' - \frac{1}{2}k_{\bot}^2) [ (k_{\bot}^2+k_zk_z') sin(\varphi-\varphi')
+ (k_z-k_z')k_{\bot}cos(\varphi-\varphi')] 
\label{D3}
\end{eqnarray}
The angle $\varphi'$ is defined by Eq. (\ref{13}) with ${\bm k}$ replaced
by ${\bm k}'$. 

The expression for $G_Q$ given by Eqs.
(\ref{111},\ref{D1},\ref{D2},\ref{D3}) holds at any value of $Q^2$. For the
evaluation of the quadrupole moment
\begin{eqnarray}
&&Q_d = \frac{\sqrt{2}}{ m_d} \lim_{Q \rightarrow 0}  \frac{1}{ Q^2}  
[ {\cal{J}}^{+}_{0,0} - {\cal{J}}^{+}_{1,1} ] = \nonumber\\
&&\lim_{Q \rightarrow 0}  \frac{2}{ Q^2} \int\nolimits 
\left[ \frac{\omega(k)\xi'}{\omega (k')\xi} \right] ^{1/2} 
\left[ (f_e^{is} - f_m^{is}) bk_{\bot} 
\frac {(b k_{\bot} E - am H)}{a^2m^2 + b^2k_{\bot}^2} - f_e^{is} \, E \right] 
\: d{\vec k}' 
\label{121}
\end{eqnarray}
an expansion of $[ {\cal{J}}^{+}_{0,0} - {\cal{J}}^{+}_{1,1} ]$ at the second
order in $\kappa$ is needed. 

Let us note that at the first order in $\kappa$ one has
\begin{eqnarray}
\varphi' - \varphi = \frac{k_{\bot} \kappa}{\xi (\omega (k) + m)}
\label{131}
\end{eqnarray}
and, as a consequence, the quantity $H$ is of the first order in $\kappa$
\begin{equation}
H = \kappa H_1 + {\cal {O}}(\kappa^2)
\label{D5}
\end{equation}
with
\begin{eqnarray}
&&H_1 = \frac{k_{\bot}}{2\xi} \left\{ \frac{1}{\omega (k) + m} \; \cdot \right.
\nonumber\\
&&\left[ - [\chi_0(k)]^2 + \frac {(k_{\bot}^2 - 2 k_z^2)}{2 k^2} 3 \varphi_2(k)
\left( 2 \sqrt{2} \varphi_0(k) + \varphi_2(k) + 3 \varphi_2(k)
\frac{\omega (k)(\omega (k) + m)}{k^2} \right) \right] - \nonumber\\
&&\left. \frac{9 \omega (k)}{\sqrt{2}k^2} \left[ \varphi_2(k) \chi_0(k)
\frac {(k_{\bot}^2 - k_z^2)}{k^2} + \frac{k_z^2}{k} 
\left( \varphi_0(k) \frac{\partial \varphi_2(k)}{\partial k} - 
\varphi_2(k) \frac{\partial \varphi_0(k)}{\partial k} \right) \right] \right\}
\label{D6}
\end{eqnarray}
Since $b$ is also of the first order in $\kappa$ (see Eq. (\ref{19'})), in
Eq. (\ref{121}) one can take $a = 2$ and disregard $b^2k_{\bot}^2$ with respect
to $a^2m^2$  in the limit $Q^2 \rightarrow 0$. As a result one has
\begin{eqnarray}
&&Q_d = \lim_{Q \rightarrow 0}  \frac{2}{Q^2}  
\int\nolimits  \left[ \frac{\omega(k)\xi'}{\omega (k')\xi} \right] ^{1/2} 
\left[ \frac{Q^2}{m_d^2} (f_e^{is} - f_m^{is}) k_{\bot} 
\frac {( k_{\bot} E - m \xi H_1)}{4m^2 \xi^2} - f_e^{is} E \right] d{\vec k}' =
\nonumber\\
&&= Q_{d1} + Q_{d2}
\label{151} 
\end{eqnarray}
where 
\begin{eqnarray}
&&Q_{d1} = \frac{2}{m_d^2}  \int\nolimits  [f_e^{is}(0) - f_m^{is}(0)] k_{\bot} 
\frac {( k_{\bot} E_0 - m \xi H_1)}{4m^2 \xi^2} \; d{\vec k} 
\label{151'}
\\
&&Q_{d2} = - \lim_{Q \rightarrow 0}  \frac{2}{Q^2}  \int\nolimits 
\left[ \frac{\omega(k)\xi'}{\omega (k')\xi} \right] ^{1/2}
f_e^{is} \, E \: d{\vec k}'
\label{151''} 
\end{eqnarray}
In the integral of Eq. (\ref{151'}) each quantity has been evaluated at
$Q^2=0$ (i.e., $\xi = \xi'$, $k_z = k_z'$) and
\begin{eqnarray}
&&E_0 = E(Q^2=0) = (\frac{1}{2}k_{\bot}^2 - k_z^2)
\frac{3\varphi_2(k)(2 \sqrt{2}\varphi_0(k) + \varphi_2(k))}{2 k^{2}} \; .
\label{D7}
\end{eqnarray}

To evaluate $Q_d$ we need an expansion of the integral in Eq. (\ref{151''}) up
to the second order in $Q$. By using the expansions of $\xi$ and $k_z$ up to the
second order in $\kappa$ given in Eq. (\ref{20'}), one obtains
\begin{eqnarray}
&&\left[ \frac{\omega(k)\xi'}{\omega (k')\xi} \right] ^{1/2} = 
1 + \kappa \Omega_1 + \frac{\kappa^2}{2} \Omega_2
\label{D8}
\\
&&E = E_0 + \kappa E_1 + \frac{\kappa^2}{2} E_2
\label{D9} 
\end{eqnarray}
where
\begin{eqnarray}
&&\Omega_1 = - \frac{4\xi' - 3}{2\xi'} \, , \quad
\Omega_2 = \frac{16\xi^{'2} - 36\xi' + 21}{4\xi^{'2}}
\label{D10}\\
\nonumber\\
&&E_1 = \frac {3 \omega (k')  k_z'} {\sqrt{2} \xi' k^{'3}} 
\left[ \frac {3 k_{\bot}^2}{k'}  \varphi_2(k') \left(\varphi_0(k') +
\frac{\varphi_2(k')}{2\sqrt{2}} \right) - \right.
\nonumber\\
&&\left.
\left( \frac {k_{\bot}^2}{2} - k_z^{'2} \right)    
\left( \varphi_0(k') \frac{\partial \varphi_2(k')}{\partial k'} +  
\varphi_2(k') \frac{\partial \varphi_0(k')}{\partial k'} +
\frac{\varphi_2(k')}{\sqrt{2}} \frac{\partial \varphi_2(k')}{\partial k'}
\right) \right]  
\label{D10'}\\
\nonumber\\
&&E_2 = \frac{k_{\bot}^2}{k^{'2}\xi ^{'2} (\omega (k') + m)^2} 
\left\{ \frac { k^{'2} [\chi_0(k')]^2}{2} - 3 \varphi_2(k')
\left( \frac {k_{\bot}^2}{2} - k_z^{'2} \right)    
\left( \sqrt{2} \chi_0(k') + \frac {3}{2}\varphi_2(k') \right) + \right.
\nonumber\\
&&\left. \frac {9 \omega (k')} {\sqrt{2} k'}  (\omega (k') + m)
\left[ \frac{\varphi_2(k')}{k'} \chi_0(k') (k_{\bot}^2 - k_z^{'2}) + \right.
\right. \nonumber\\
&&\left. \left. k_z^{'2} \left( \varphi_0(k') 
\frac{\partial \varphi_2(k')}{\partial k'} -   \varphi_2(k') 
\frac{\partial \varphi_0(k')}{\partial k'} \right) + \sqrt{2}
\frac{[\varphi_2(k')]^2}{k'}  
\left( \frac {k_{\bot}^2}{2} - k_z^{'2} \right) 
\right] \right\} + 
\nonumber\\
&&\frac{3 \omega (k')  k'_z}{\sqrt{2} \xi ^{'2} k^{'3}} (4 \xi' - 3)
\left\{ \left( \frac {k_{\bot}^2}{2} - k_z^{'2}\right)
\left[ \varphi_2(k') \frac{\partial \varphi_0(k')}{\partial k'} +
\varphi_0(k') \frac{\partial \varphi_2(k')}{\partial k'} +
\frac{\varphi_2(k')}{\sqrt{2}} \frac{\partial \varphi_2(k')}{\partial k'} 
\right] - \right.
\nonumber\\
&&\left. \frac {3 k_{\bot}^2}{k'} \varphi_2(k') 
\left( \varphi_0(k') + \frac{\varphi_2(k')}{2\sqrt{2}} \right) \right\} +
\nonumber\\
&&\frac{3 [\omega (k')]^2}{\sqrt{2} \xi ^{'2} k^{'4}}
\left\{ \left( \frac {k_{\bot}^2}{2} - k_z^{'2} \right)
\left[ \varphi_2(k') \left( \frac {k_{\bot}^2}{k'}
\frac{\partial \varphi_0(k')}{\partial k'} +
k_z^{'2} \frac{\partial^2 \varphi_0(k')}{\partial k^{'2}} \right) +
\right. \right.
\nonumber\\
&&\left. \left.
\varphi_0(k') \left( \frac {k_{\bot}^2}{k'}
\frac{\partial \varphi_2(k')}{\partial k'} +  k_z^{'2} 
\frac{\partial^2 \varphi_2(k')}{\partial k^{'2}} \right) + 
\frac {k_z^{'2}}{\sqrt{2}}\varphi_2(k')  
\frac{\partial^2 \varphi_2(k')}{\partial k^{'2}} \right] + \right. 
\nonumber\\
&&\left. \frac {k_{\bot}^2} {k'} \left[  \frac {3} {k'} \varphi_0(k')
\varphi_2(k') (3 k_z^{'2} - k_{\bot}^2) - 6 k_z^{'2}  \varphi_0(k')
\frac{\partial \varphi_2(k')}{\partial k'} +
\frac {3\sqrt{2} k_z^{'2}} {k'} [\varphi_2(k')]^2 + \right. \right.
\nonumber\\
&&\left. \left. \sqrt{2} \left( \frac {k_{\bot}^2}{4} - 2 k_z^{'2} \right) 
\varphi_2(k') \frac{\partial \varphi_2(k')}{\partial k'}
\right] \right\} \; .
\label{D11}
\end{eqnarray}
Furthermore, because of Eq. (\ref{119}), one has
\begin{eqnarray}
&&f_e^{is}((p_1' - p_1)^2) = 1 + Q^2 
\left[ \frac{df_e^{is}((p_1' - p_1)^2)}{d (Q^2)} \right]_{Q^2=0} = 
1 - \frac{(r^{is}_e)^2}{3}\frac{2\kappa^2(m^2+k_{\bot}^2)}{\xi\xi'}
\label{D12}
\end{eqnarray}
where
\begin{eqnarray}
(r^{is}_e)^2 = 6 \left[ \frac{df_e^{is}((p_1' - p_1)^2)}{d ((p_1' - p_1)^2)}
\right]_{Q^2=0} = r_{ep}^2 + r_{en}^2
\label{D13}
\end{eqnarray}
is the sum of the squares of the proton and neutron charge mean square radii
(let us recall that $(p_1' - p_1)^2 \le 0$). Then, since only the second order
terms in the expansion of the integral in Eq. (\ref{151''}) can give a
contribution to $Q_d$, one obtains \begin{eqnarray}
Q_{d2} = -  \frac{1}{4m_d^2}  \int\nolimits 
\left[ \Omega_2 E_0 + E_2 + 2 \Omega_1  E_1 - 
4 E_0 \frac{(r^{is}_e)^2}{3}\frac{(m^2 + k_{\bot}^2)}{\xi ^2} \right] 
d{\vec k} 
\label{D14} 
\end{eqnarray}
where each quantity has to be evaluated at $Q^2=0$.

 

\pagebreak
\begin{center}

TABLE CAPTION
\end{center}

Table I. Magnetic moment (in nuclear magnetons) and quadrupole moment for the
deuteron, corresponding to different $N-N$ interactions; $\mu _d^{NR}$
and $Q_d^{NR}$ are the nonrelativistic results, $\mu _d$ (LPS) and $Q_d$ (LPS)
our results; $P_D$ is the $D$-state percentage, and $\eta = A_D/A_S$ the
symptotic normalization ratio (this table is taken from Ref. \cite{LPS3}, a
part from the results for the $Nijmegen2$ interaction).

\begin{table}
       
\begin{center} 
\begin{tabular}{|c|c|c|c|c|c|c|}
\hline
Interaction & $P_D$  & $\eta$ & $\mu _d^{NR}$ &
$\mu_d$ (LPS) & $Q_d^{NR}$ $fm^{2}$  & $Q_d$ (LPS) $fm^{2}$\\
\hline 
         &   &   &  &  &  &\\
Exp  &  & 0.0256(4) \cite{eta} &  &   0.857406(1) \cite{Lind}&  &
0.2859(3) \cite{Rosa}\\ 
RSC \cite{RSC} & 6.47 & 0.0262   & 0.8429   & 0.8611 & 0.2796 & 0.2852 \\ 
Av14 \cite{AV14} & 6.08 & 0.0265 & 0.8451   & 0.8608 & 0.2860 & 0.2907 \\
Paris \cite{Par} & 5.77 & 0.0261 & 0.8469   & 0.8632 & 0.2793 & 0.2841 \\
Av18 \cite{AV18} & 5.76 & 0.0250  & 0.8470  & 0.8635 & 0.2696 & 0.2744 \\
Nijm93 \cite{Nij} & 5.75 & 0.0252 & 0.8470  & 0.8629 & 0.2706 & 0.2750 \\
RSC93 \cite{Nij} & 5.70 & 0.0251 & 0.8473   & 0.8637 & 0.2703 & 0.2750 \\
Nijm1 \cite{Nij} & 5.66 & 0.0253 & 0.8475   & 0.8622 & 0.2719 & 0.2758 \\
Nijm2 \cite{Nij} & 5.64 & 0.0252 & 0.8477   & 0.8652 & 0.2707 & 0.2756 \\
CD-Bonn \cite{CDB} & 4.83 & 0.0255 & 0.8523 & 0.8670 & 0.2696 & 0.2729 \\ 
\hline 
\end{tabular} 
\end{center} 
\label{table2} 
\end{table}

\vspace {3cm}
Table I  F.M. LEV, E. PACE, G. SALM\`E 
\newpage
\begin{center}

FIGURE CAPTIONS
\end{center}

FIG. 1. (a) Deuteron form factor $A(Q^2)$ obtained using the
$RSC$ $N-N$ interaction \cite{RSC} and the Gari-Kr\"{u}mpelmann nucleon form
factors \cite{Gari}. Solid line: full result of our approach with the
Poincar\'e covariant current operator. Dashed line: the argument of the nucleon
form factors, $(p_1' - p_1)^2)$, is  replaced  by $-Q^2$. Long-dashed line:
nonrelativistic result obtained with exact relativistic relations
between deuteron form factors and current matrix elements, within the
Breit reference frame where  ${\hat{q}} = {{\bm e}_z}$
\cite{Gourdin}, but with nonrelativistic expressions for the matrix elements
evaluated in impulse approximation \cite {Lomon}. 
Experimental data are from Ref. \cite{Galster} (open squares),
Ref. \cite{Saclay} (triangles), Ref. \cite{SLAC}
(diamonds), Ref. \cite{JLABA} (full dots) and \cite{JLABC} (open dots).
(b) The same as in (a), but
for $B(Q^2)$. Experimental data are from Ref. \cite{Buchanan} (open dots), Ref.
\cite{Ganichot} (open squares), Ref. \cite{Sac-Auffret} (full diamonds), Ref.
\cite{Cramer} (triangles), Ref. \cite{SLAC-Bosted} (full squares), and 
\cite{JLABAB} (open diamonds). 
(c) The same as in (a), but for $T_{20}(Q^2)$.  Experimental data are
from  Ref. \cite{Bates84} (open dots), Ref. \cite{Novo85} (full triangles), Ref.
\cite{Novo90} (open triangles), Ref. \cite{The} (full dots), Ref. \cite{Ferro}
(open squares), Ref. \cite{NIKHEF} (full squares), and Ref. \cite{JLABCT}
(diamonds). \\

FIG. 2. (a) As in Fig. 1 (a), but for the reduced form factor 
$A(Q^2)/(G_D^2 \cdot F)$  with  $G_D = (1 + Q^2/0.71)^{-2}$ and $F = (1 +
Q^2/0.1)^{-2.5}$.  (b) As in Fig. 1 (b), but for the reduced form factor
$\Gamma_M (Q^2) / ( G_D^2 \cdot F_1 )$ 
with $\Gamma_M (Q^2) = [G_M (Q^2) m_p / (\mu_d m_d)]^2$ and
$F_1 = (1 + Q^2/0.1)^{-3}$. Experimental data are as in Fig. 1. \\

FIG. 3. (a) The deuteron form factor $A(Q^2)$ obtained using our
Poincar\'e covariant current operator, different $N-N$ 
interactions and the nucleon form factors by H\"{o}hler et al.
\cite{Hoehler}. Solid line: RSC interaction \cite{RSC}; dashed line: $AV18$
interaction \cite{AV18}; dot-dashed line: $Nijmegen1$ interaction;
long-dashed line: $Nijmegen2$ interaction; short-dashed line: $Nijmegen 93$
interaction \cite{Nij}; dotted line: $CD-Bonn$ interaction \cite{CDB}.
Actually the $Nijmegen 93$ result is very similar to the $AV18$ one and is
not reported in this figure. (b) The same as in (a), but for $B(Q^2)$.  (c) The
same as in (a), but for $T_{20}(Q^2)$.  Experimental data are as in Fig. 1.
\\

FIG. 4. (a) As in Fig. 3 (a), but for the reduced form factor 
$A(Q^2)/(G_D^2 \cdot F)$. (b) As in (a), but at low
$Q^2$. (c) As in Fig. 3 (b), but for the reduced form factor
$\Gamma_M (Q^2) / ( G_D^2 \cdot F_1 )$. The $Nijmegen1$ result
is very similar to the $CD-Bonn$ one and is not reported in this figure.
Experimental data are as in Fig. 1
\\

FIG. 5. (a) The position of the minimum of $B(Q^2)$, and 
(b) the position of the second zero of $T_{20}(Q^2)$, corresponding to 
the Gari-Kr\"{u}mpelmann nucleon form factors \cite{Gari}, vs the
nonrelativistic S-state kinetic energy for the deuteron for different realistic
interactions.
\\

FIG. 6. (a) The reduced deuteron form factor $A(Q^2)/(G_D^2 \cdot F)$
obtained with the $Nijmegen2$ interaction for different nucleon form factor
models. Solid line: nucleon f.f. of Ref. \cite{Mergell}; dashed line: nucleon
f.f. of  Ref. \cite{Hoehler}; dotted line: nucleon f.f. of Ref. \cite{Gari}.
(b) As in (a), but for the reduced form factor
$\Gamma_M (Q^2) / ( G_D^2 \cdot F_1 )$. 
(c) As in (a), but for $T_{20}(Q^2)$.
Experimental data are as in Fig. 1.
\\

\newpage
%

\psfig{bbllx=0mm,bblly=170mm,bburx=150mm,bbury=280mm,file=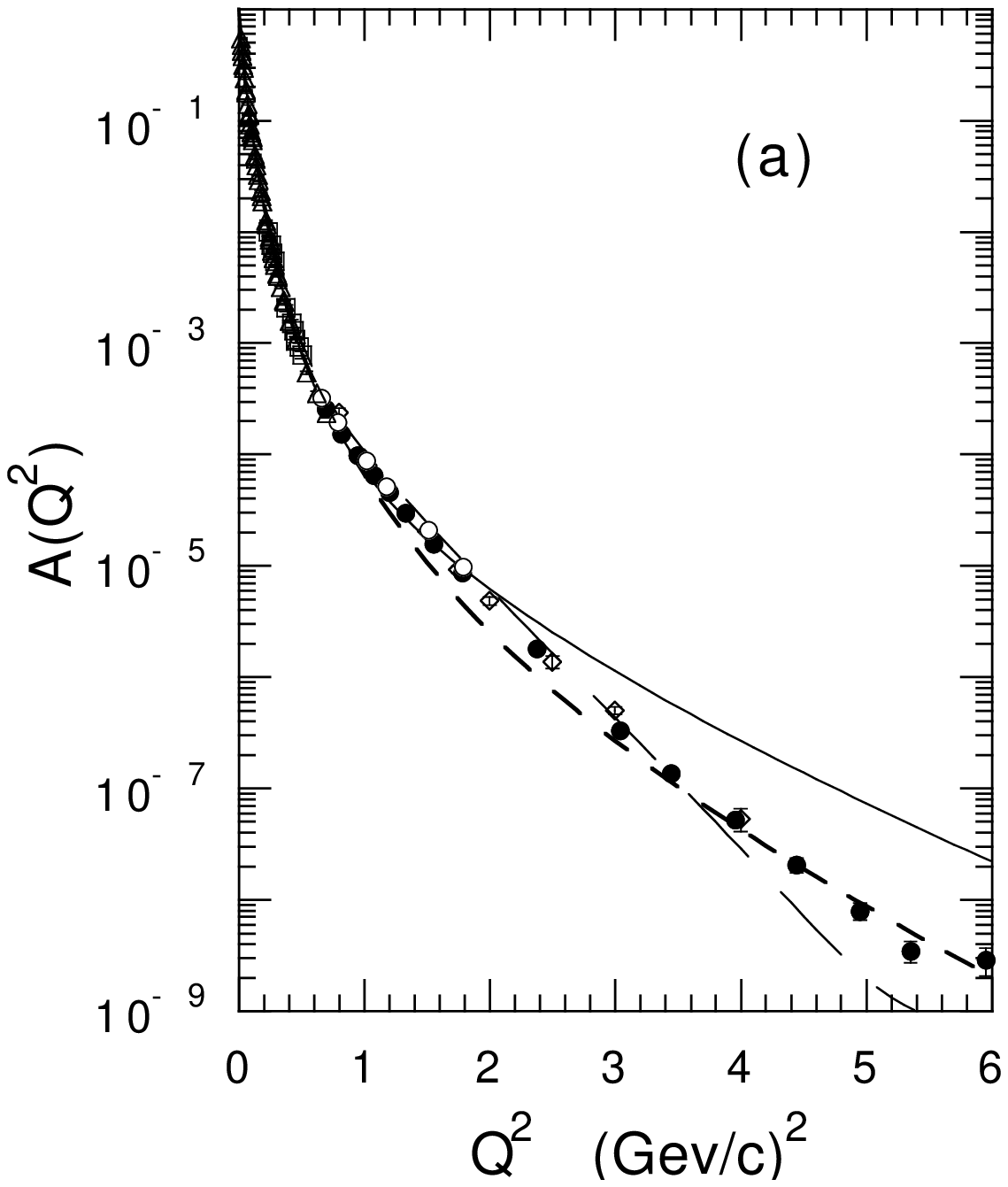}

\vspace {3cm}
Fig. 1a  F.M. LEV, E. PACE, G. SALM\`E 
%

\newpage
%

\psfig{bbllx=0mm,bblly=170mm,bburx=150mm,bbury=280mm,file=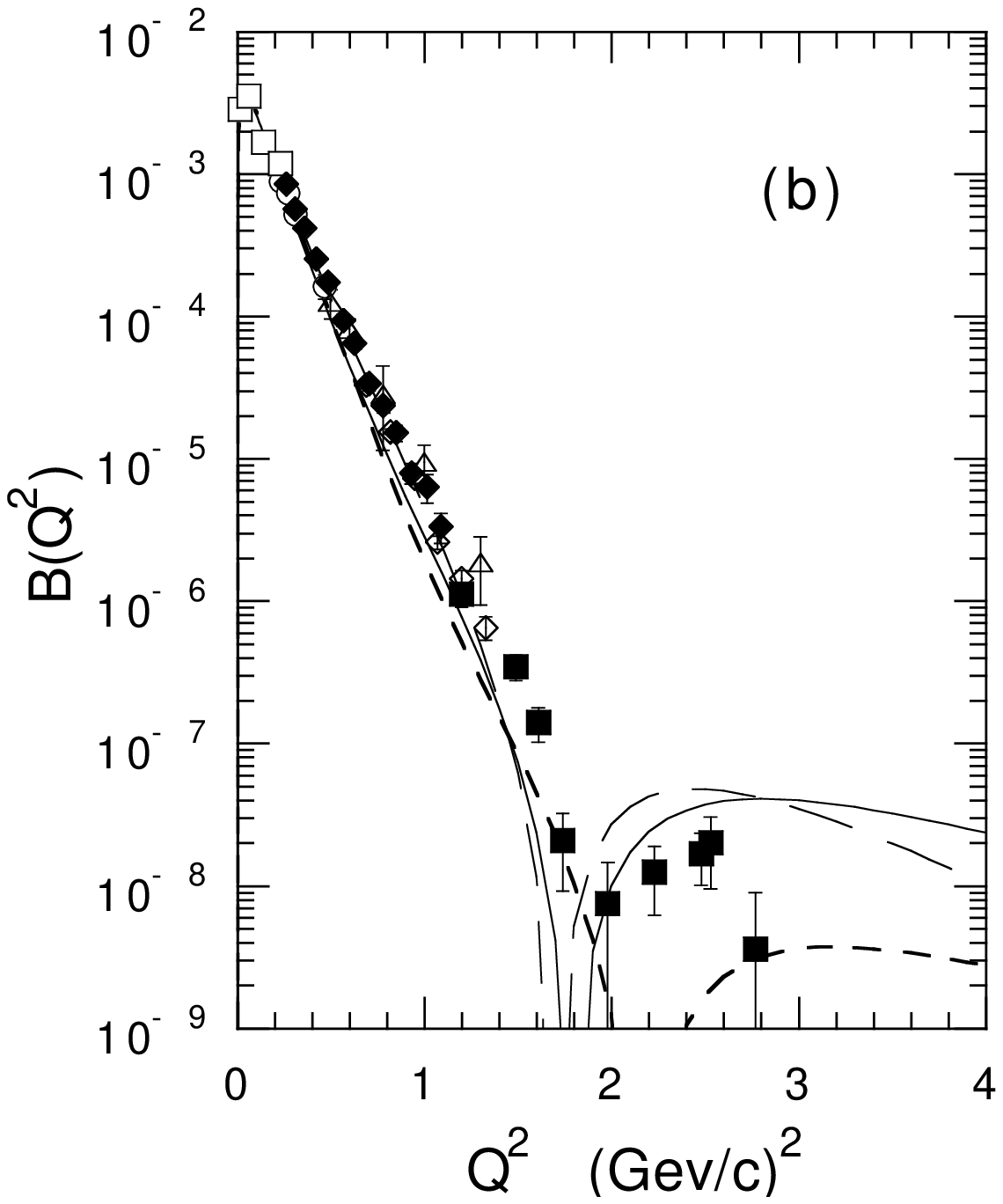}

\vspace {3cm} 
Fig. 1b  F.M. LEV, E. PACE, G. SALM\`E   

\newpage
%

\psfig{bbllx=0mm,bblly=170mm,bburx=150mm,bbury=280mm,file=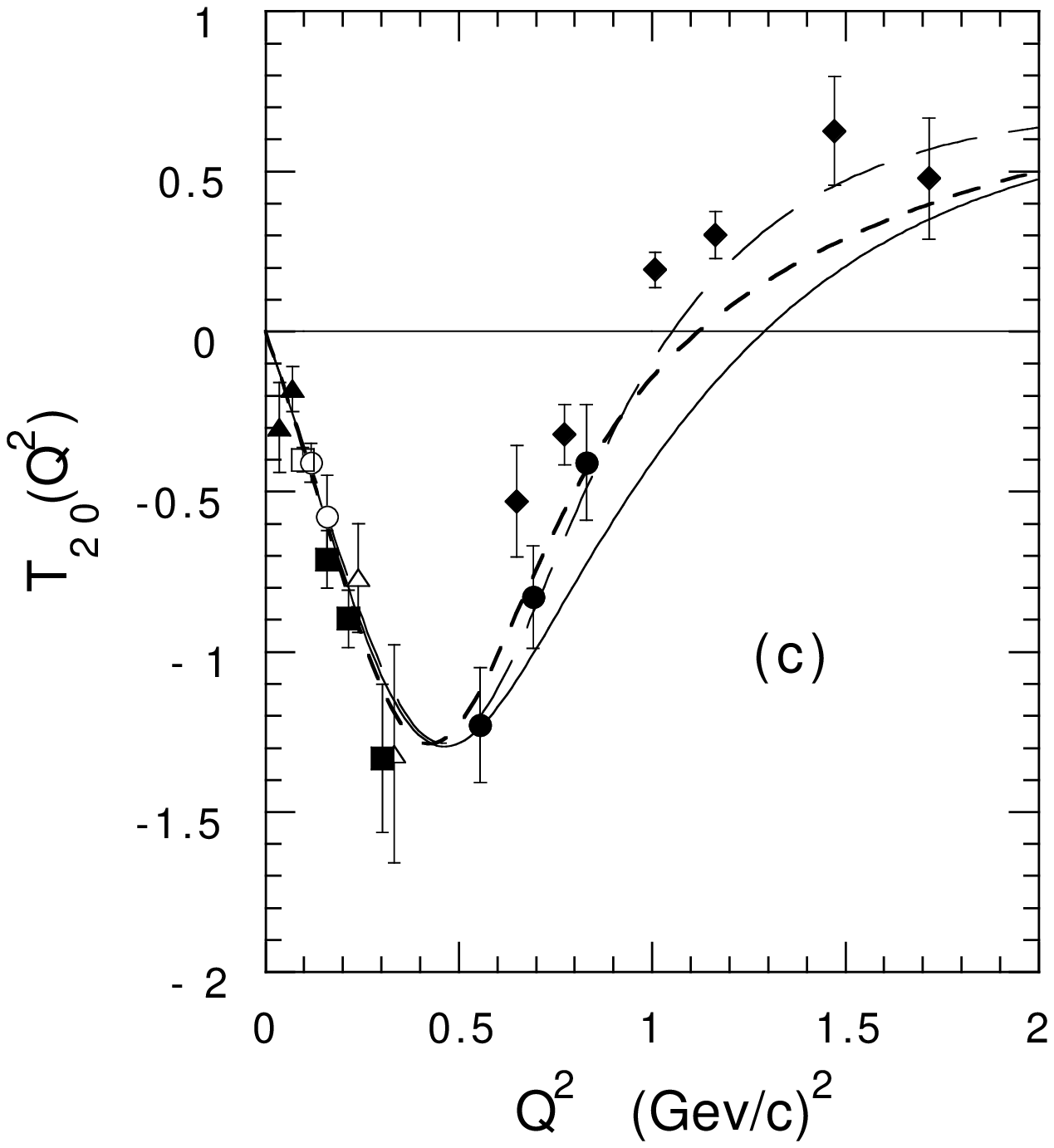}

\vspace {3cm} 
Fig. 1c  F.M. LEV, E. PACE, G. SALM\`E   

\newpage
%

\psfig{bbllx=0mm,bblly=170mm,bburx=150mm,bbury=280mm,file=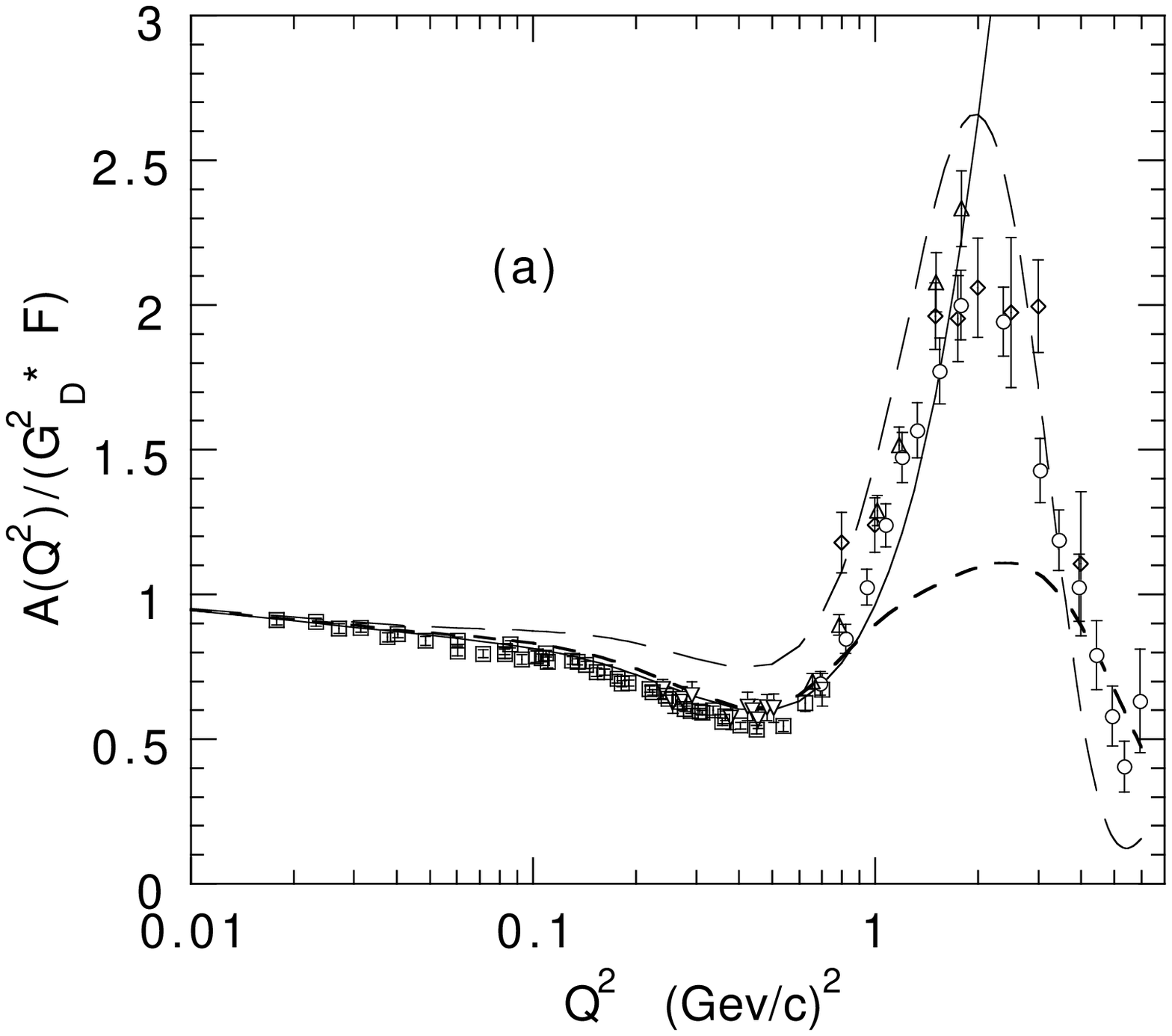}

\vspace {5cm} 
Fig. 2a  F.M. LEV, E. PACE, G. SALM\`E   

\newpage
%

\psfig{bbllx=0mm,bblly=170mm,bburx=150mm,bbury=280mm,file=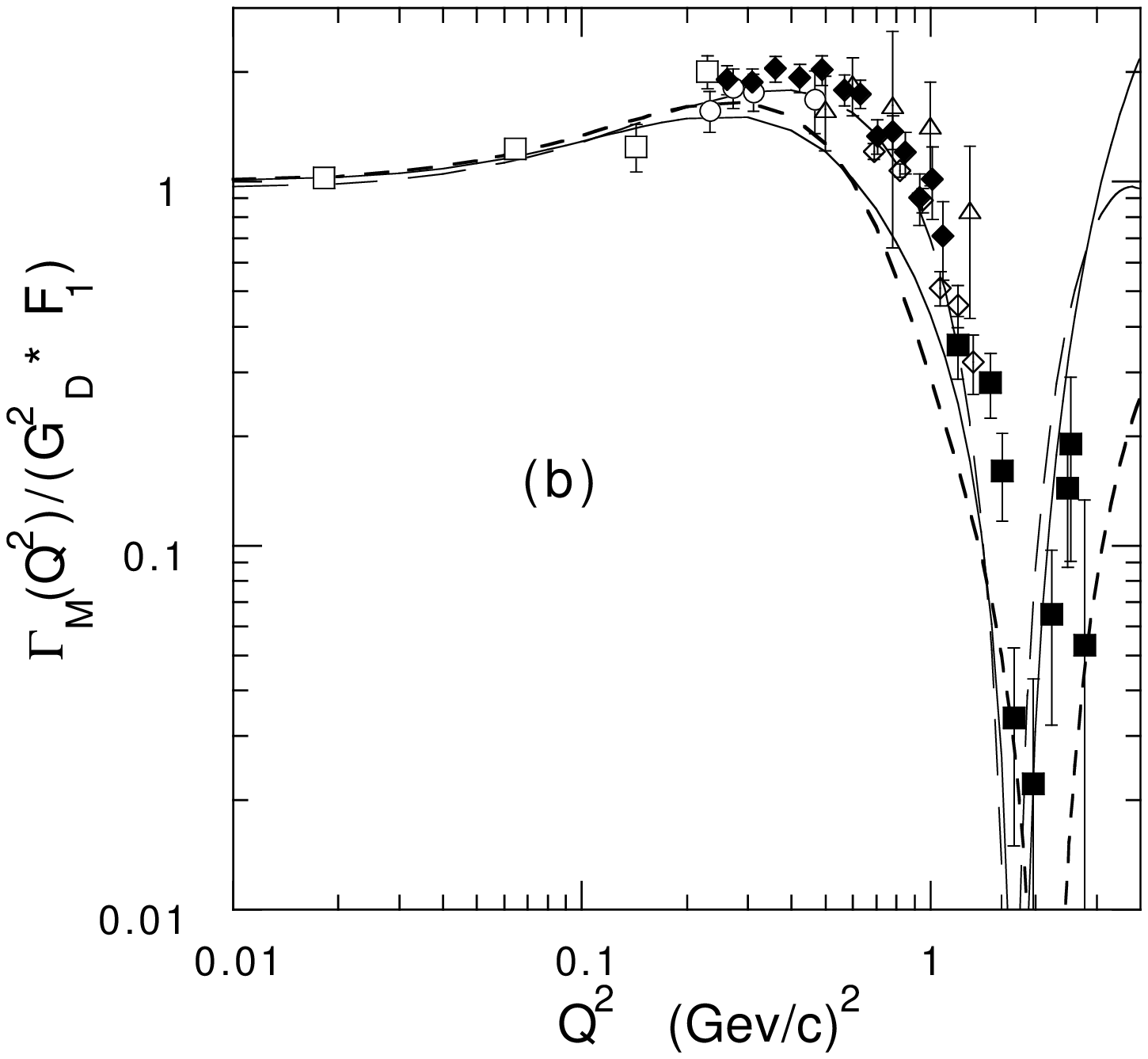}

\vspace {3cm} 
Fig. 2b  F.M. LEV, E. PACE, G. SALM\`E   

\newpage
%
\psfig{bbllx=0mm,bblly=170mm,bburx=150mm,bbury=280mm,file=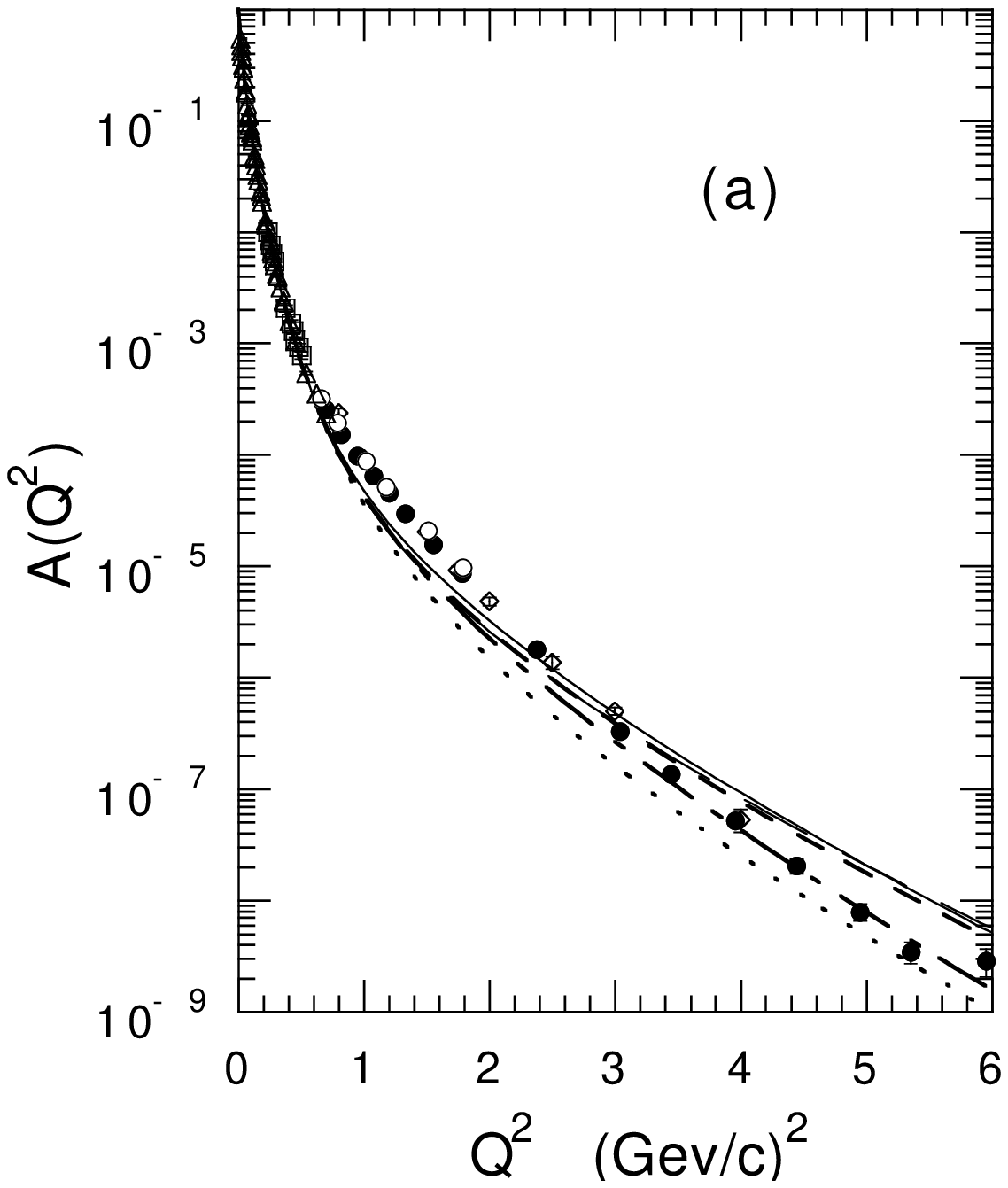}

\vspace {3cm} 
Fig. 3a  F.M. LEV, E. PACE, G. SALM\`E   

\newpage
%

\psfig{bbllx=0mm,bblly=170mm,bburx=150mm,bbury=280mm,file=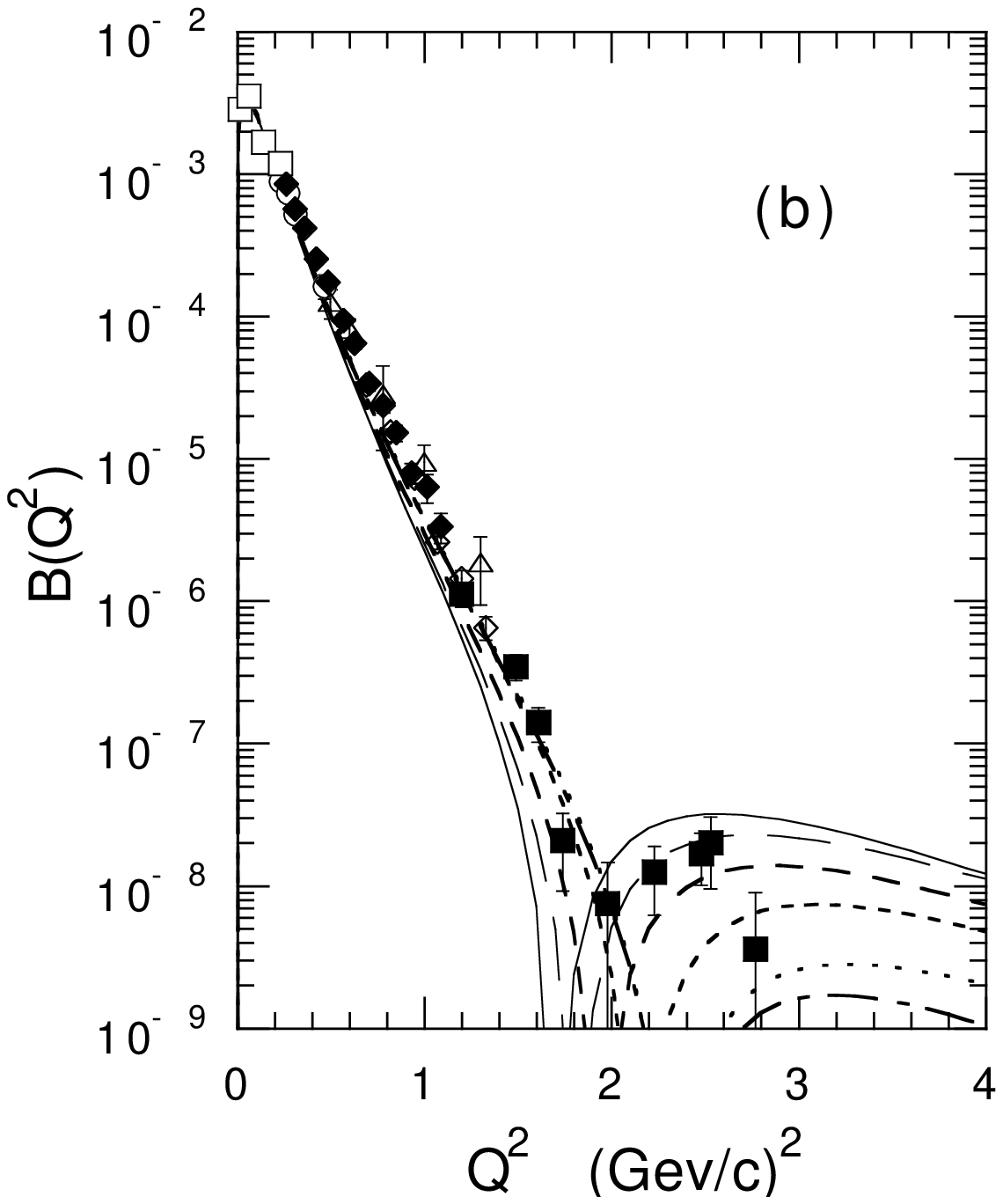}

\vspace {3cm} 
Fig. 3b  F.M. LEV, E. PACE, G. SALM\`E   

\newpage
%

\psfig{bbllx=0mm,bblly=170mm,bburx=150mm,bbury=280mm,file=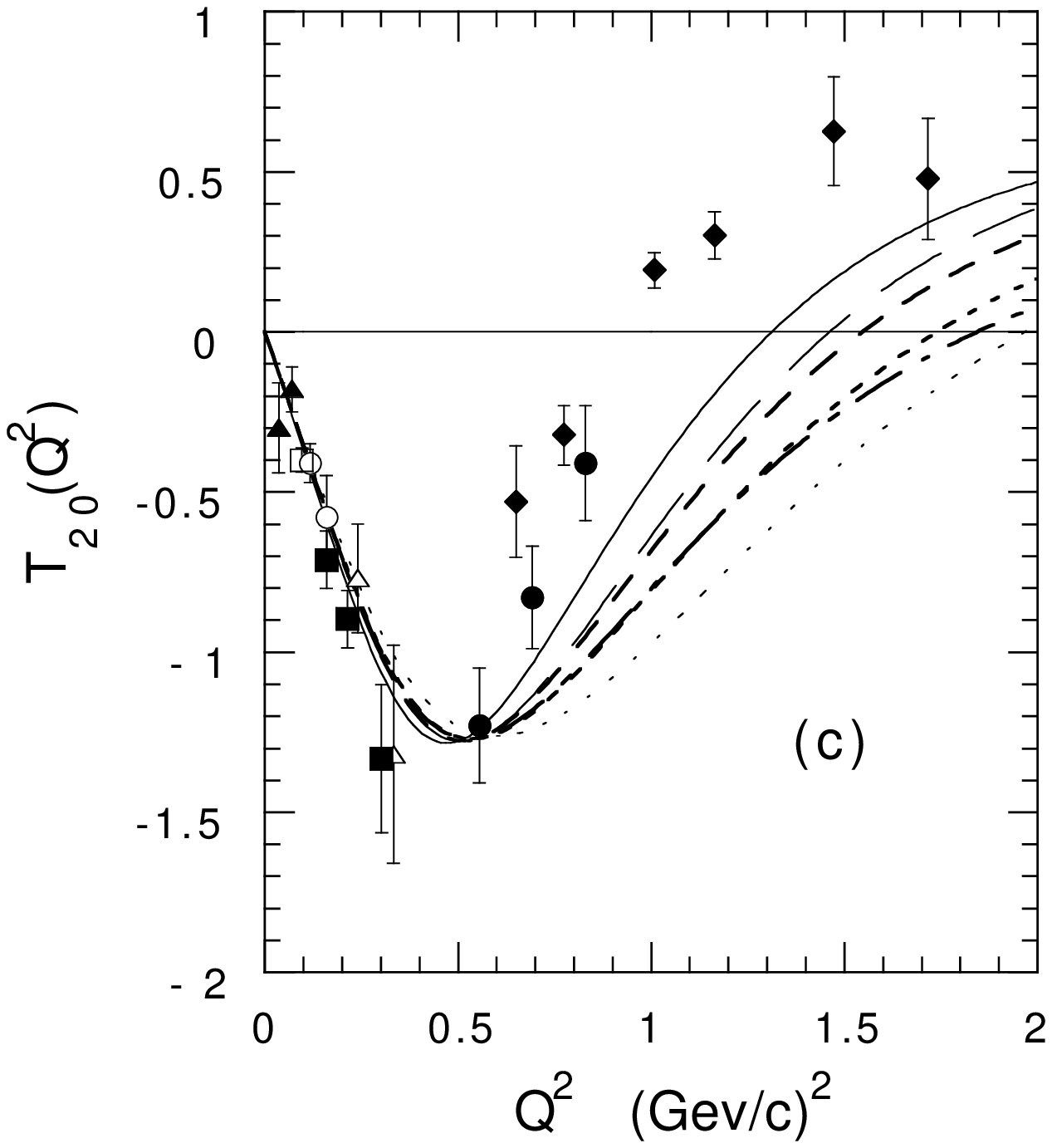}

\vspace {3cm} 
Fig. 3c  F.M. LEV, E. PACE, G. SALM\`E   

\newpage
%

\psfig{bbllx=0mm,bblly=170mm,bburx=150mm,bbury=280mm,file=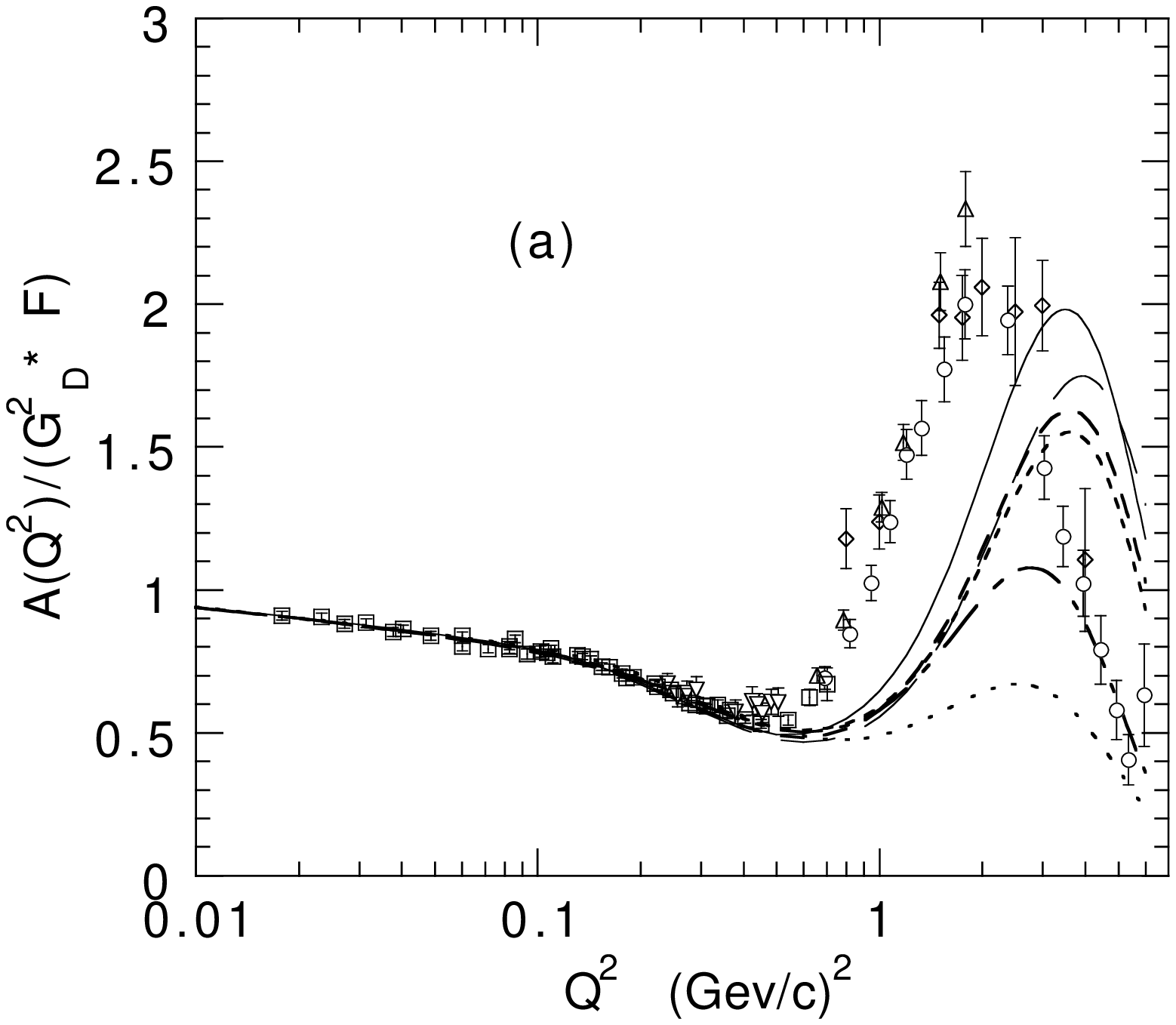}

\vspace {5cm} 
Fig. 4a  F.M. LEV, E. PACE, G. SALM\`E  
 
\newpage
%

\psfig{bbllx=0mm,bblly=170mm,bburx=150mm,bbury=280mm,file=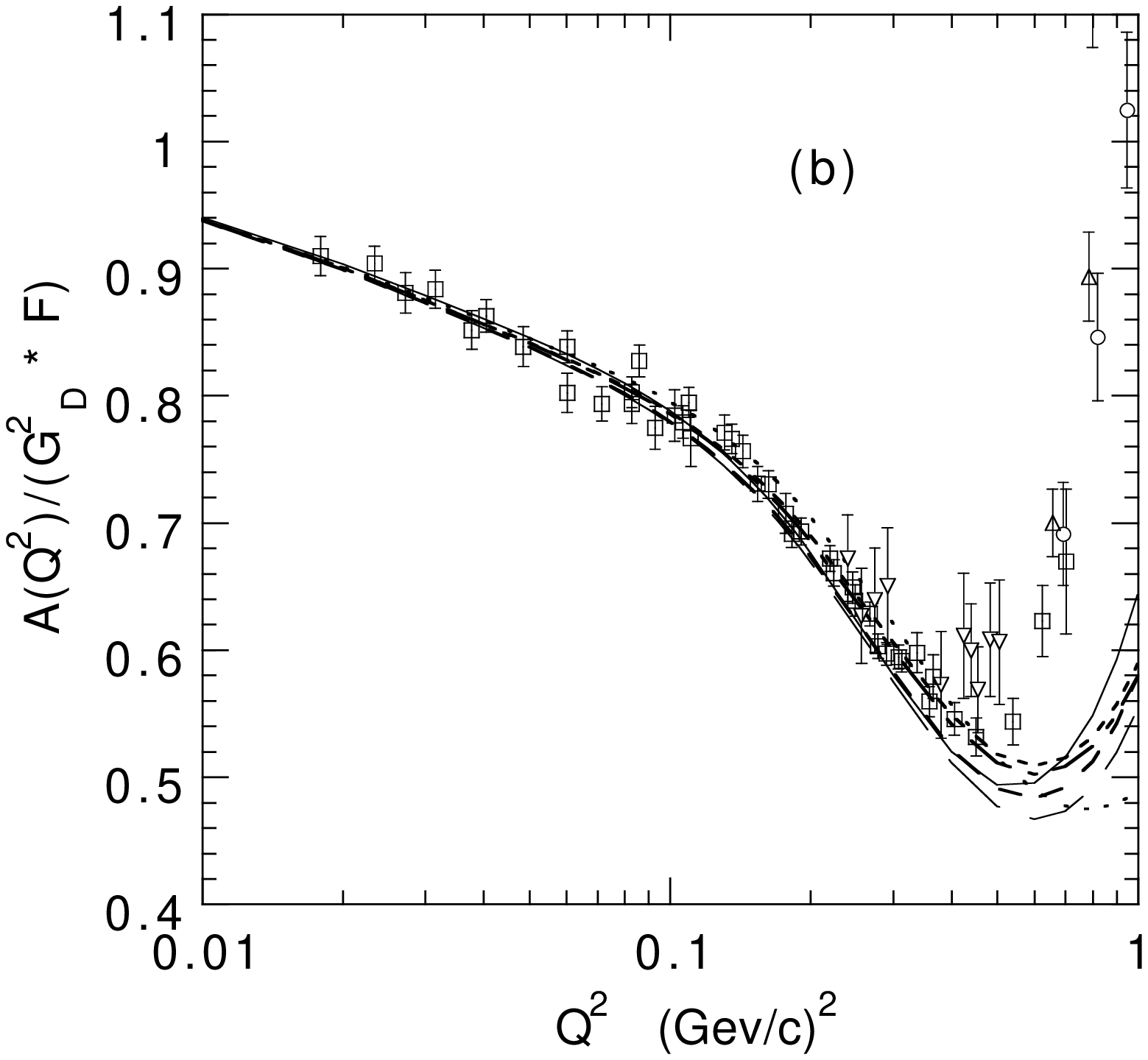}

\vspace {3cm} 
Fig. 4b  F.M. LEV, E. PACE, G. SALM\`E  
 
\newpage
%

\psfig{bbllx=0mm,bblly=170mm,bburx=150mm,bbury=280mm,file=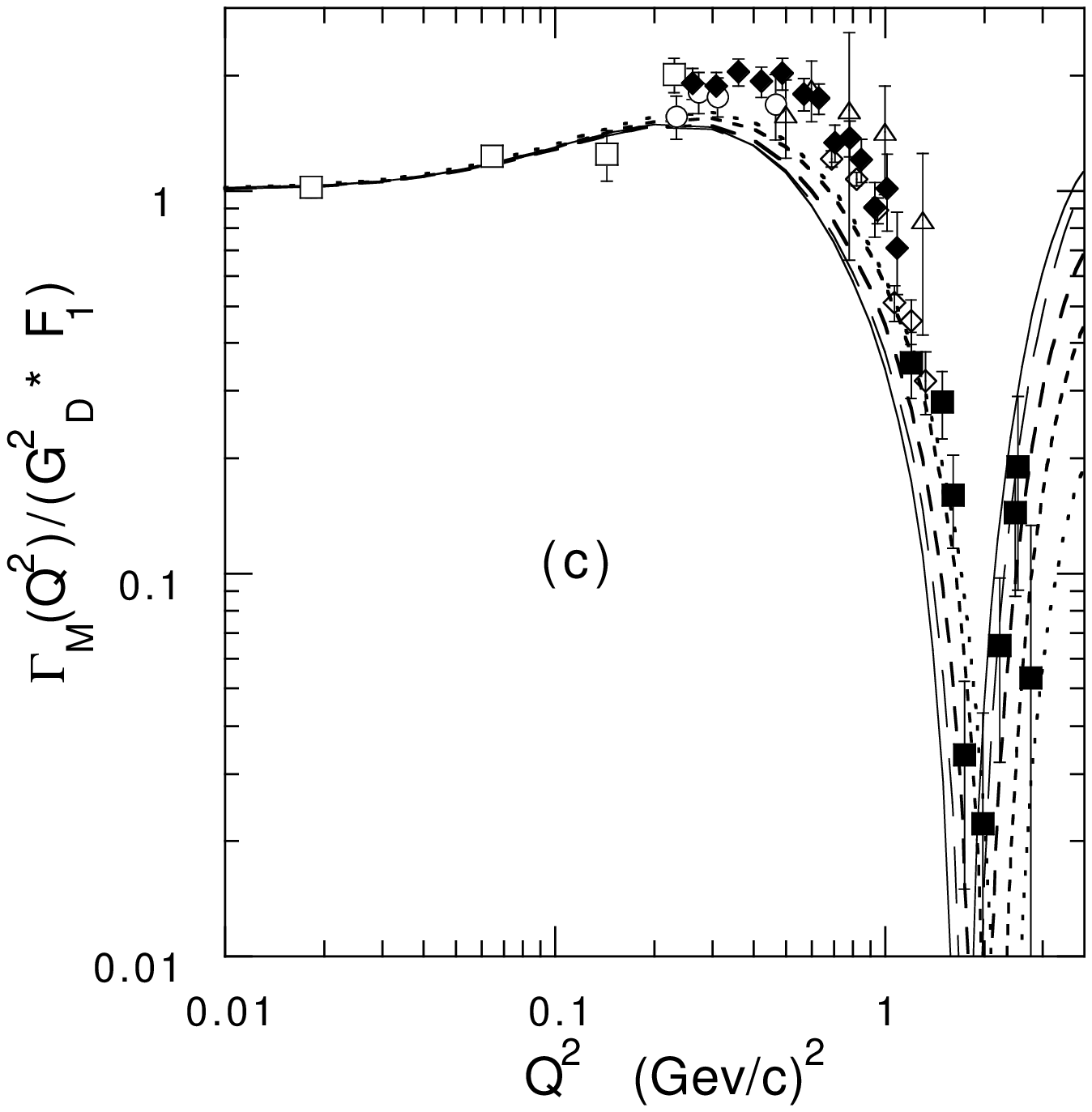}

\vspace {3cm} 
Fig. 4c  F.M. LEV, E. PACE, G. SALM\`E 
  
\newpage
%

\psfig{bbllx=0mm,bblly=170mm,bburx=150mm,bbury=280mm,file=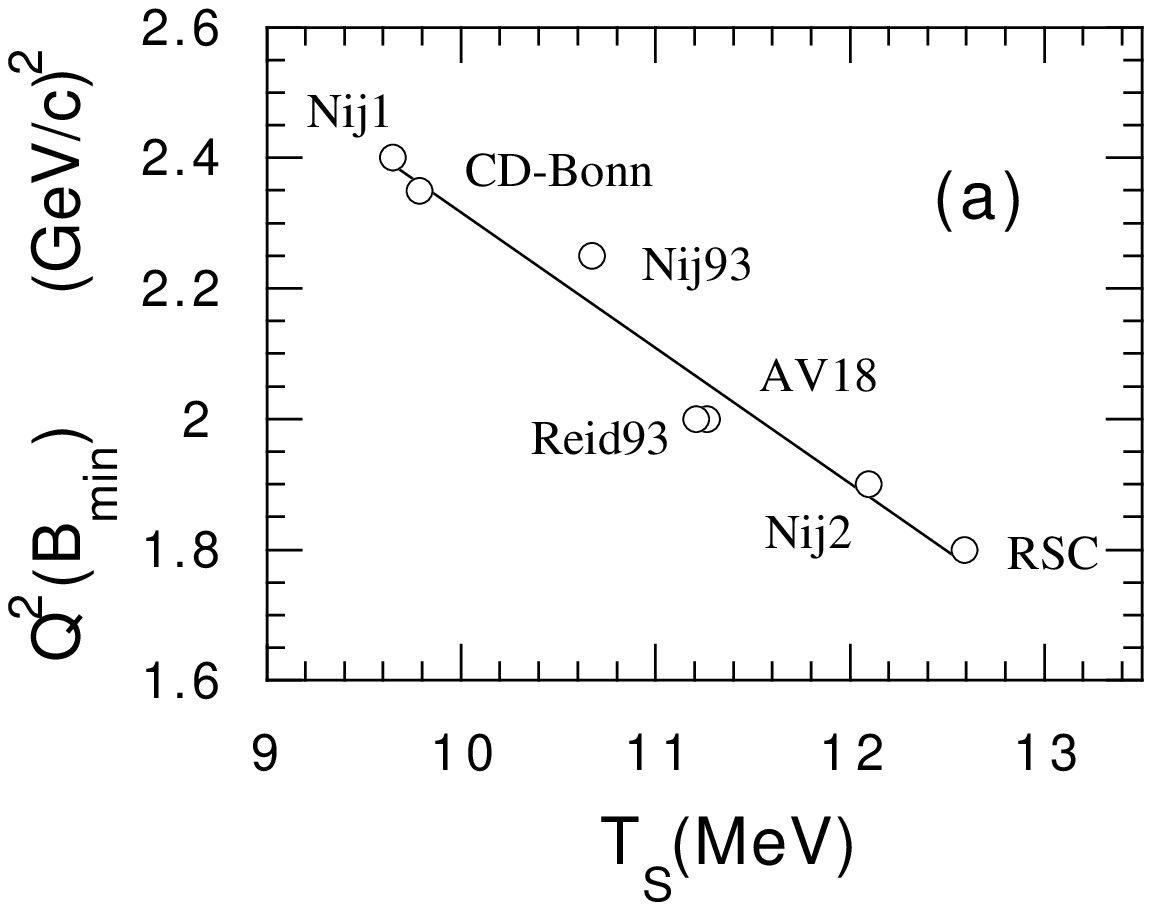}

\vspace {3cm} 
Fig. 5a  F.M. LEV, E. PACE, G. SALM\`E 
  
\newpage
%

\psfig{bbllx=0mm,bblly=170mm,bburx=150mm,bbury=280mm,file=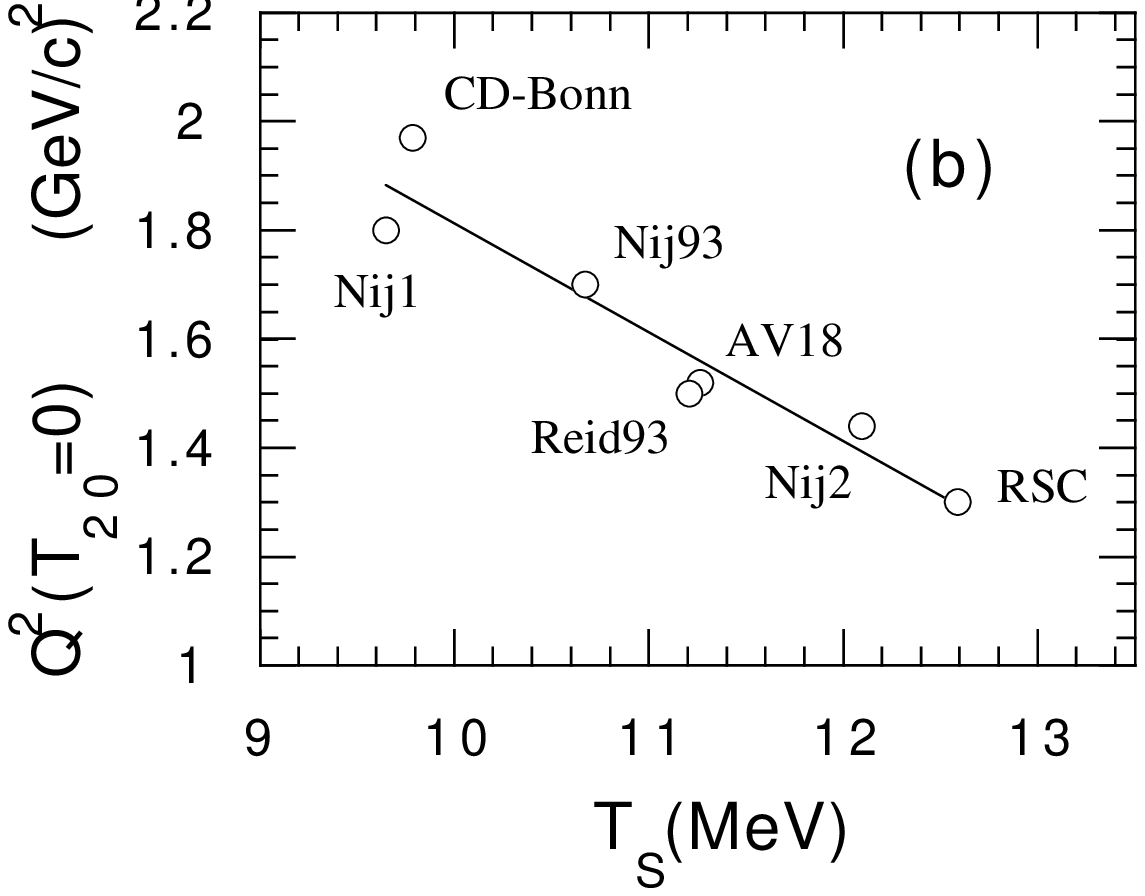}

\vspace {3cm} 
Fig. 5b  F.M. LEV, E. PACE, G. SALM\`E   

\newpage
%

\psfig{bbllx=0mm,bblly=170mm,bburx=150mm,bbury=280mm,file=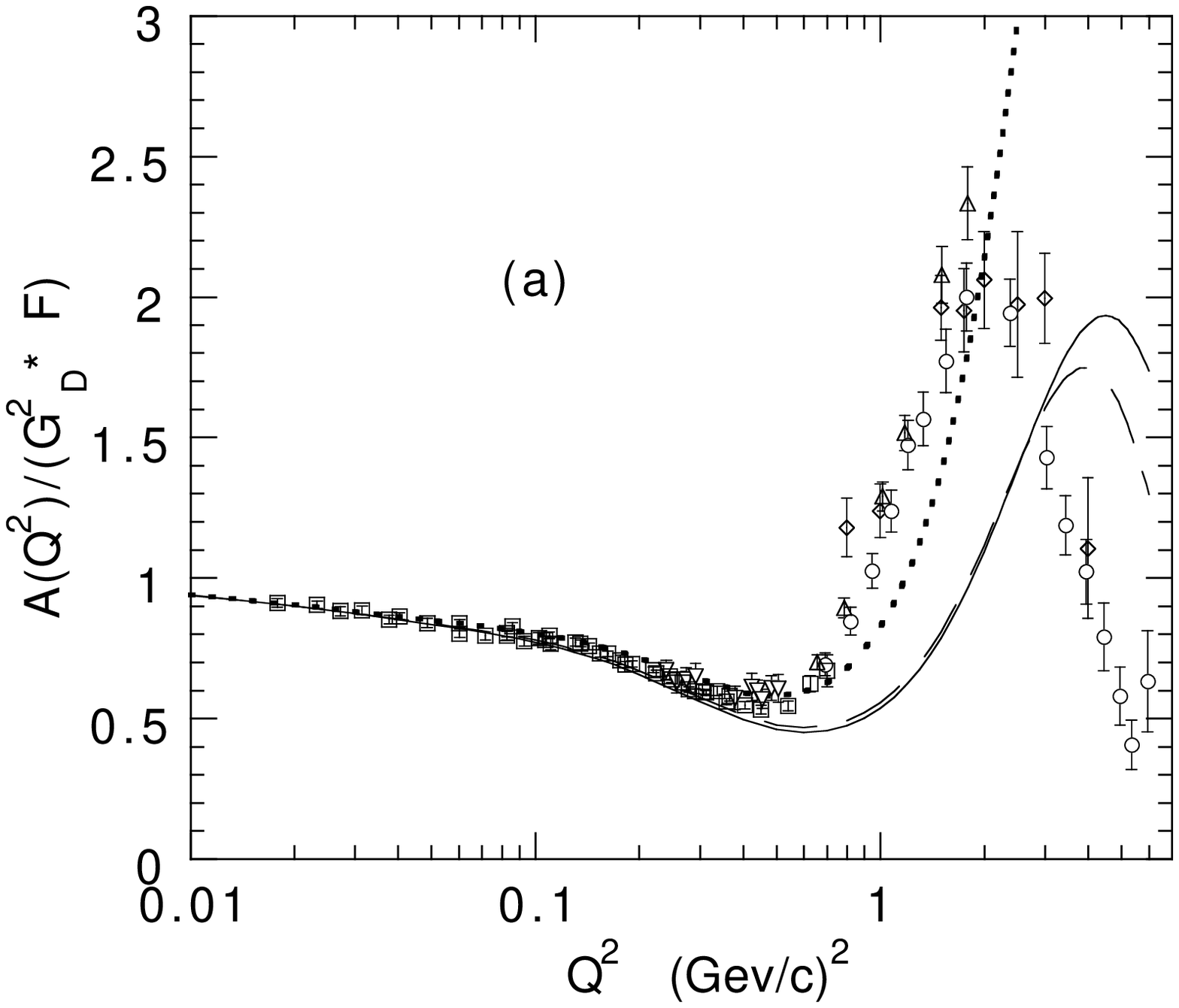}

\vspace {5cm} 

Fig. 6a  F.M. LEV, E. PACE, G. SALM\`E   

\newpage
%

\psfig{bbllx=0mm,bblly=170mm,bburx=150mm,bbury=280mm,file=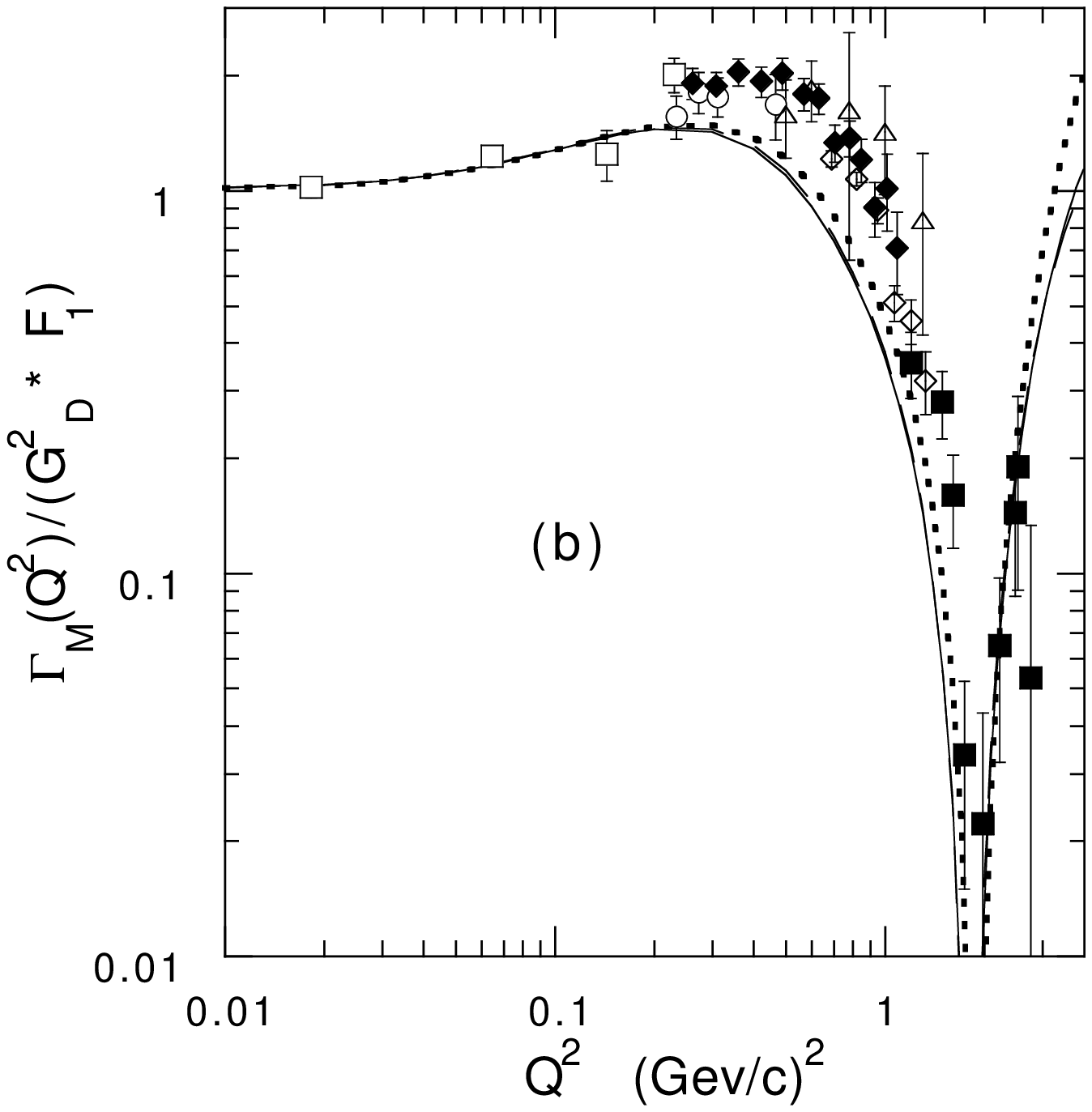}

\vspace {3cm} 
Fig. 6b  F.M. LEV, E. PACE, G. SALM\`E   

\newpage
%

\psfig{bbllx=0mm,bblly=170mm,bburx=150mm,bbury=280mm,file=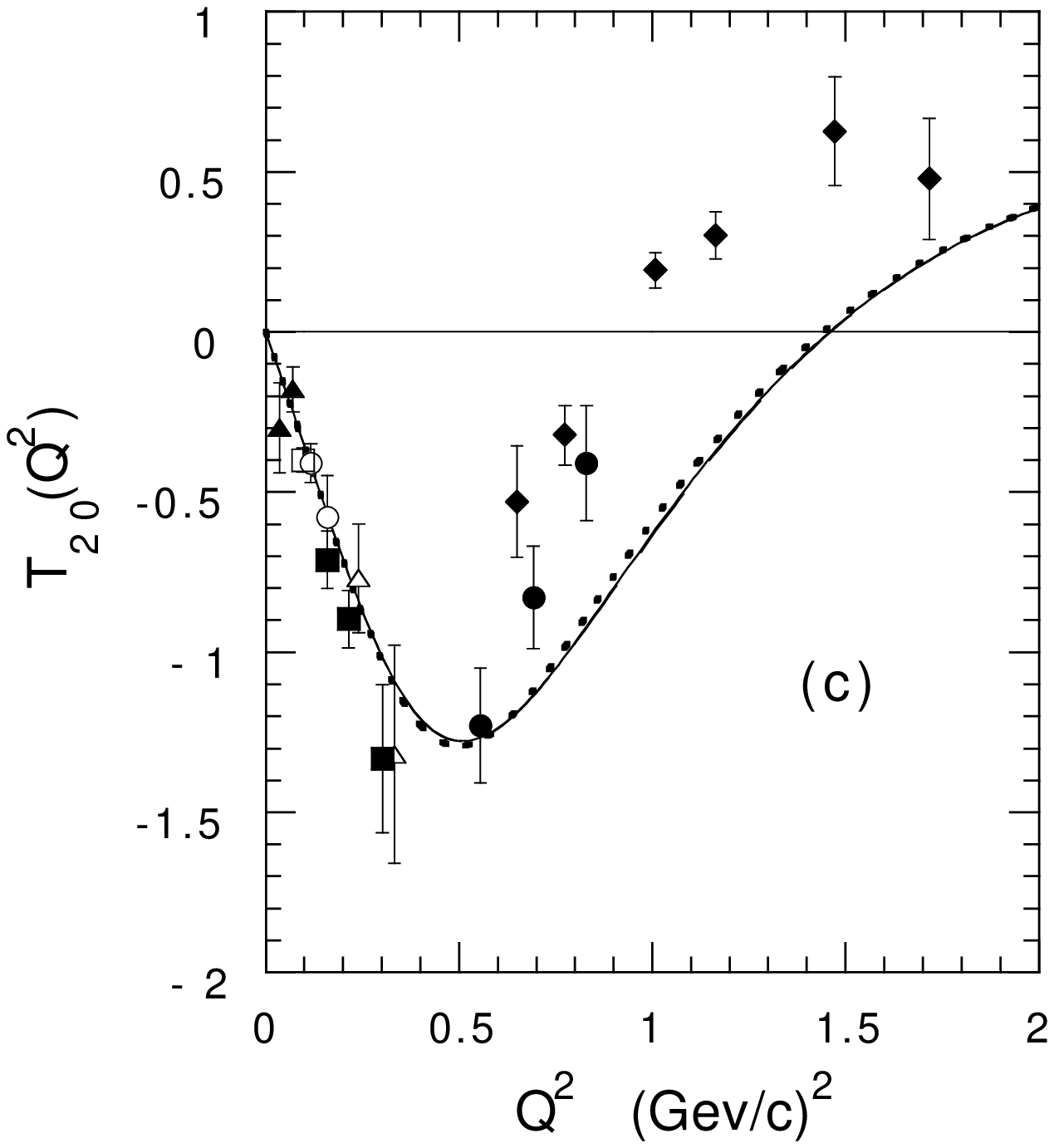}

\vspace {3cm} 
Fig. 6c  F.M. LEV, E. PACE, G. SALM\`E   

\end{document}